\pdfoutput=1
\documentclass[a4paper,fleqn]{cas-sc}

\usepackage[numbers]{natbib}

\usepackage{graphicx}
\usepackage{color,soul}
\usepackage{array}
\usepackage{amsmath}
\usepackage{amsfonts}
\usepackage{amssymb}
\usepackage{psfrag}
\usepackage{epstopdf}
\usepackage{subfig}
\usepackage{booktabs}
\usepackage{hyperref}
\usepackage{latexsym}
\usepackage{kantlipsum}
\usepackage{balance}
\usepackage{lscape}
\usepackage{rotating}
\usepackage{multicol}
\usepackage{multirow,bigdelim}
\usepackage{color, colortbl}
\usepackage{threeparttable}
\usepackage{float}

\usepackage{tikz}
\usetikzlibrary{calc,fit,shapes,arrows,positioning,chains,arrows.meta}
\usepackage{makeshape}
\usepackage{caption}
\usepackage{floatflt}

\newcommand{\parder}[2]{\frac{\partial #1}{\partial #2}}

\newcommand{\secder}[2]{\frac{\partial^2 #1}{\partial #2^2}}

\newcommand{\bzero}{\mathbf{0}}

\newcommand{\eval}[2][\right]{\relax\ifx#1\right\relax \left.\fi#2#1\rvert}

\newcommand{\ba}{\boldsymbol{a}}

\newcommand{\be}{\boldsymbol{e}}
\newcommand{\bff}{\boldsymbol{f}}
\newcommand{\bn}{\boldsymbol{n}}
\newcommand{\bt}{\boldsymbol{t}}

\newcommand{\bv}{\boldsymbol{v}}
\newcommand{\bw}{\boldsymbol{w}}
\newcommand{\bx}{\boldsymbol{x}}
\newcommand{\by}{\boldsymbol{y}}

\newcommand{\bE}{\boldsymbol{E}}
\newcommand{\bF}{\boldsymbol{F}}

\newcommand{\del}{\boldsymbol{\nabla}}
\newcommand{\bcross}{\boldsymbol{\times}}

\makeatletter
\newcommand{\Rmnum}[1]{\expandafter\@slowromancap\romannumeral #1@}
\makeatother
\newcommand{\bA}{\boldsymbol{A}}
\newcommand{\bz}{\boldsymbol{z}}

\newcommand{\bK}{\boldsymbol{K}}

\newcommand{\bimu}{\frac{1}{\mu}}

\newcommand{\ksqbimu}{\frac{k^2}{\mu}}

\newcommand{\br}{\boldsymbol{r}}

\newcommand{\bH}{\boldsymbol{H}}

\newcommand{\bpsi}{\boldsymbol{\Psi}}
\newcommand{\bpsihat}{\hat{\bpsi}}

\newcommand{\delpsi}{\del\psi}
\newcommand{\psihat}{\hat{\psi}}

\newcommand{\bAhat}{\hat{\bA}}

\newcommand{\emikx}{e^{-ik|\boldsymbol{x}|}}

\newcommand{\modx}{|\boldsymbol{x}|}

\newcommand{\bAbar}{\bar{\bA}}

\newcommand{\psibar}{\bar{\psi}}

\newcommand{\bAbarhat}{\hat{\bAbar}}

\newcommand{\psibarhat}{\hat{\psibar}}

\newcommand{\bi}{\boldsymbol{i}}
\newcommand{\bj}{\boldsymbol{j}}
\newcommand{\bk}{\boldsymbol{k}}

\def\tsc#1{\csdef{#1}{\textsc{\lowercase{#1}}\xspace}}
\tsc{WGM}
\tsc{QE}
\tsc{EP}
\tsc{PMS}
\tsc{BEC}
\tsc{DE}

\begin{document}
\let\WriteBookmarks\relax
\def\floatpagepagefraction{1}
\def\textpagefraction{.001}
\shorttitle{}
\shortauthors{Durgarao Kamireddy, Sreekanth Karanam, Arup Nandy}

\title [mode = title]{A novel implementation of symmetric boundary condition in harmonic and transient analysis of electromagnetic wave propagation}





\author[]{Durgarao Kamireddy}
\cormark[1]
\ead{durga176103010@iitg.ac.in}


\address[]{Department of Mechanical Engineering, Indian Institute of Technology Guwahati, Guwahati 781039, India}
\author[]{Sreekanth Karanam}
\ead{skaranam@iitg.ac.in}
\author[]{Arup Nandy}
\ead{arupn@iitg.ac.in}

\cortext[cor1]{Corresponding author}







\begin{abstract}
 While doing electromagnetic analysis using FEM (Finite element method), if we can implement the underlying symmetric nature of the problem, there will be significant reduction in the computational cost. Symmetric nature of the problem can be identified from the given physical loading and boundary conditions of the problem. But for electromagnetic analysis in potential formulation, it is not very straight forward to implement the symmetric boundary condition. In the present work, a novel implementation of symmetric boundary condition in potential formulation within nodal framework, has been demonstrated. The implementation has been carried out in both electromagnetic harmonic and transient analysis for a wide range of radiation and scattering problems. A significant reduction in computational cost is achieved as compared to the existing formulation.
\end{abstract}




\maketitle

\section{Introduction}

 In the field of computational electromagnetics, FEM has been implemented to simulate a wide range of application problems in antenna radiations, wave guide transmissions, scattering by conducting and dielectric bodies etc.~\cite{antenna,jin}. While solving electromagnetic radiation and scattering problems using FEM, the infinite exterior domain is required to be truncated at some finite radius as our computational domain should be finite. To mimic the infinite nature, some appropriate absorbing boundary condition (ABC) is imposed on the truncation surface~\cite{harmonic}. This ABC is determined based on Sommerfeld radiation condition~\cite{em2}. In~\cite{em2,kane,stu,web}, various second- and higher-order ABCs were implemented. In~\cite{em2,kane,stu,web}, authors modeled the interior computational domain with edge FEM and the truncation surface with Boundary Integral (BI) method. Even though this technique is more accurate than traditional ABCs, a key disadvantage is the generation of densely populated matrix by BI elements, which results in higher memory requirement and computational time.

     For certain electromagnetic problems, from the given loading and boundary conditions, we can identify some plane of symmetry. In such situations, we can consider half of the domain on any side of the plane for our computational analysis. Additionally, some suitable boundary conditions are required to implement on that plane of symmetry. Altogether, it reduces the overall computational cost. It is not very straightforward to implement this symmetric condition in potential formulation~\cite{eigen}. In this work, we have proposed some novel technique to implement this symmetric boundary condition in potential formulation in the nodal framework. For the problems in exterior domain, suitable ABC is implemented on the truncation boundary.

     Being inspired from the Wilcox asymptotic expansion for electric field in exterior domain~\cite{Mittra:wilcox}, A. Nandy et. al. proposed an amplitude formulation~\cite{harmonic} in nodal FEM, using potentials, to carry out harmonic electromagnetic analysis, for solving electromagnetic radiation and scattering problems in exterior domain. In the current work, we have also extended our novel implementation of symmetric boundary condition in the Amplitude formulation~\cite{harmonic}.  
 
   A wide range of numerical methods are available to solve Maxwell's electromagnetic wave equation in time domain. They can be classified in two broad categories such as Finite Difference Time Domain (FDTD) method and Time Domain FEM (TDFEM) method. FDTD, introduced by Yee~\cite{pile2015}, gained popularity due to its ease in implementation of the numerical integration. TDFEM~\cite{Chen2009,Member1994,Wong1995,Movahhedi2007,Jiao2002a,Lee2002,Tsai2002} soon replaced FDTD as it could be easily generalized to complex geometries. In TDFEM, two approaches were followed to solve for electromagnetic field variables. In first approach~\cite{Jiao2002a,Lee2002,Tsai2002}, second order electromagnetic wave equation was solved for either electric or magnetic field. Then from post processing, other field was obtained. In the other approach~\cite{Chen2009,Movahhedi2007,Wong1995}, two coupled first order Maxwell's equations were solved for both field variables simultaneously. In~\cite{Faghihi2008, Jiao2002a,Liu2001, Qiu2007}, Finite Element Boundary Integral (FEBI) method was developed for exterior domain problems. In transient electromagnetic analysis, different time stepping strategies such as leapfrog strategy~\cite{Rieben2005,Rodrigue2001}, central difference method~\cite{jiao2003,Jiao2002a,Liu2001} were used for temporal discretization. Stability of the algorithm is an important factor in simulating transient electromagnetic analysis. In~\cite{Nandy2018}, the time-stepping strategy for TDFEM was unconditionally stable from an energy perspective. In this current manuscript, the novel implementation of the symmetric boundary condition has been carried out in both harmonic and transient electromagnetic analysis.

\par     The remaining article is organized as follows: in section~\ref{math_formulation}, we have briefly presented the mathematical and FEM formulation for both harmonic and transient electromagnetic analysis. We have presented the implementation of symmetric boundary condition in potential formulation in section~\ref{method2}. Thereafter, we have selected various electromagnetic radiation and scattering examples in both harmonic (section~\ref{numerical_examples_harmonic}) and transient (section~\ref{numerical_examples_transient}) analysis in such a way that there exist some plane of symmetry in those problems. We have implemented our novel method in those problems to demonstrate the computational efficacy of the proposed method as compared to the existing formulations. 

\section{Mathematical Formulation}
\label{math_formulation}
In this section, after brief presentation of the mathematical and FEM formulation for harmonic and transient electromagnetic analysis using potential formulation, we elaborately present implementation of the symmetric boundary condition.

    Maxwell's electromagnetic wave equation in terms of electric field $\bE$~\cite{grif} can be written as
\begin{equation} \label{eqmaxwell1}
	\frac{\epsilon_r}{c^2}\secder{\bE}{t}+\mu_0\parder{\bj}{t}+\del\bcross\left (\frac{1}{\mu_r}\del\bcross\bE\right)=\bzero,
\end{equation}

where $c=1/\sqrt{\epsilon_0\mu_0}$ is the speed of light and $\bj$ is the current density. Relative permittivity, $\epsilon_r$ and relative permeability, $\mu_r$ are given as $\epsilon_r=\frac{\epsilon}{\epsilon_0}$ and $\mu_r=\frac{\mu}{\mu_0}$, where $\epsilon_0$ and $\mu_0$ are the electric permittivity and magnetic permeability for vacuum, and $\epsilon$ and $\mu$ are the electric permittivity and magnetic permeability of the medium, respectively. 
\subsection{Harmonic electromagnetic analysis}
\label{harmonic_formulation}
For harmonic excitation, Eq.~\ref{eqmaxwell1} can be expressed as
\begin{equation} \label{govern_har}
\del\bcross\left(\bimu\del\bcross\bE\right) - \ksqbimu\bE = -i\omega\bj,
\end{equation}
where $i=\sqrt{-1}$, $\omega$ is the excitation frequency, $k=k_0\sqrt{\mu_r\epsilon_r}$ is the wave number of the medium, and $k_0=\omega/c$ is the wave number of the vacuum. For no charge condition, the electric field is subjected to the following constraint
\begin{equation} \label{eqsomcond1}
\del\cdot (\epsilon\bE)=0.
\end{equation}

For perfectly conducting boundary ($\varGamma_e$), $\bE\bcross\bn=0$, whereas for other part ($\varGamma_h$), $\bH\bcross\bn$ is prescribed. Assuming no surface current at material discontinuity, both $\bE\bcross\bn$ and $\bH\bcross\bn$ must be continuous across the material interface~\cite{jin}. In Potential formulation, electric field $\bE$ is replaced as $\bA+\delpsi$. Then, the governing differential equation given by Eq.~\eqref{govern_har}, and the constraint equation given by Eq.~\eqref{eqsomcond1}, can be written as
\begin{subequations}
	\begin{gather}
	\del\bcross\left(\bimu\del\bcross\bA\right) - \ksqbimu(\bA+\del\psi) = -i\omega\bj, \label{govern_poten} \\
	\del\cdot (\epsilon\bA)+\del\cdot (\epsilon\del\psi)=0. \label{govern_poten_div}
	\end{gather}
\end{subequations}
In the variational statements of the Eqs.~\eqref{govern_poten} and \eqref{govern_poten_div}, we have to include a regularization term corresponding to the Coulomb gauge condition $\del \cdot\bA=\text{constant}$, and the first-order absorbing boundary condition on truncated boundary ($\varGamma_\infty$)~\cite{harmonic}. Finally, the finite element equation can be expressed as,  
\begin{equation}
\begin{bmatrix} \bK_{AA} & \bK_{A\psi} \\ \bK_{\psi A} & \bK_{\psi\psi} \end{bmatrix}
\begin{bmatrix} \bAhat \\ \psihat \end{bmatrix} = \begin{bmatrix} \bF_A  \\ \bF_\psi \end{bmatrix}. \label{discreteP}
\end{equation}
Detailed derivation of the above FE equation are presented in~\cite{harmonic}.

     From the expansion theorem of Wilcox~\cite{wilcox, Mittra:wilcox}, for exterior domain electromagnetic problems, field $\bE$ can be presented as
\begin{equation} 
\bE = \frac{\emikx}{\modx}\sum_{n=0}^{\infty}\frac{\ba_n(\theta,\phi)}{\modx^n},
\end{equation}
where $\modx$ is the distance of a point from the origin, or, alternatively, the radial coordinate in the spherical coordinate system $(r,\theta,\phi)$. Being inspired from this expansion theorem, Amplitude formulation for harmonic electromagnetic analysis has been proposed in~\cite{harmonic}. Here, rapidly varying part $\frac{\emikx}{\modx}$ is separated out apriori from the potentials as

\begin{subequations}
	\begin{gather}
        \bA = \frac{1}{\modx}\bar\bA(\bx)\emikx, \\
        \psi = \frac{1}{\modx}\bar\psi(\bx)\emikx.
	\end{gather}
\end{subequations}
FEM has to capture a relatively gentle variation of $\bar\bA$ and $\bar\psi$. Finally the resulting finite element equations are given as 
\begin{equation}
\begin{bmatrix} \bK_{AA} & \bK_{A\psi} \\ \bK_{\psi A} & \bK_{\psi\psi} \end{bmatrix}
\begin{bmatrix} \bAbarhat \\ \psibarhat \end{bmatrix} = \begin{bmatrix} \bF_A  \\ \bF_\psi \end{bmatrix}. \label{discretePowf}
\end{equation}
Different terms in the above equations have been derived elaborately in~\cite{harmonic}.


\subsection{Transient electromagnetic analysis}
\label{transient_formulation}
For transient electromagnetic analysis using potential formulation~\cite{Nandy2018}, electric and magnetic fields are expressed in terms of $\bv:=\partial\bA/\partial t$ and $\bw:=\partial(\del\psi)/\partial t$ as  
\begin{align}
\bE&=-\bw-\bv, \label{eqmaxwell41d} \\
\bH&=\frac{1}{\mu}\del\bcross\bA. \label{eqmaxwell41e}
\end{align}
Then, Eq.~\ref{eqmaxwell1} and Eq.~\ref{eqsomcond1} are expressed as 
\begin{gather}
	\epsilon\left[\secder{\bA}{t}+\del \left(\secder{\psi}{t}\right)\right]+\parder{\bj}{t}+\del\bcross\left(\frac{1}{\mu}\del\bcross\bA\right)=\bzero, \label{eqmaxwell40} \\
	\del\cdot\left[\epsilon\left(\del\parder{\psi}{t}+\parder{\bA}{t}\right)\right]=0.
\end{gather}
In transient analysis choice of time stepping strategy is very crucial to conserve certain physical quantities with a view to make our numerical formulation unconditionally stable. After following the detailed derivations as depicted in~\cite{Nandy2018}, the finite element equation can be written as

\begin{equation}
\begin{bmatrix}
\bK_{AA} & \bK_{A\psi} \\
\bK_{\psi A} & \bK_{\psi\psi} \end{bmatrix}\begin{bmatrix} \bAhat_{n+1} \\ \bpsihat_{n+1} \end{bmatrix}=\begin{bmatrix}\bff_A \\ \bff_{\psi} \end{bmatrix}.
\label{fullmatAphi}
\end{equation}


\subsection{Implementation of the symmetric boundary condition using a thin patch}
\label{method2} 

\begin{figure}[pos=h!]
	\centering
	\includegraphics[width = 3.0 cm, height= 4.5 cm]{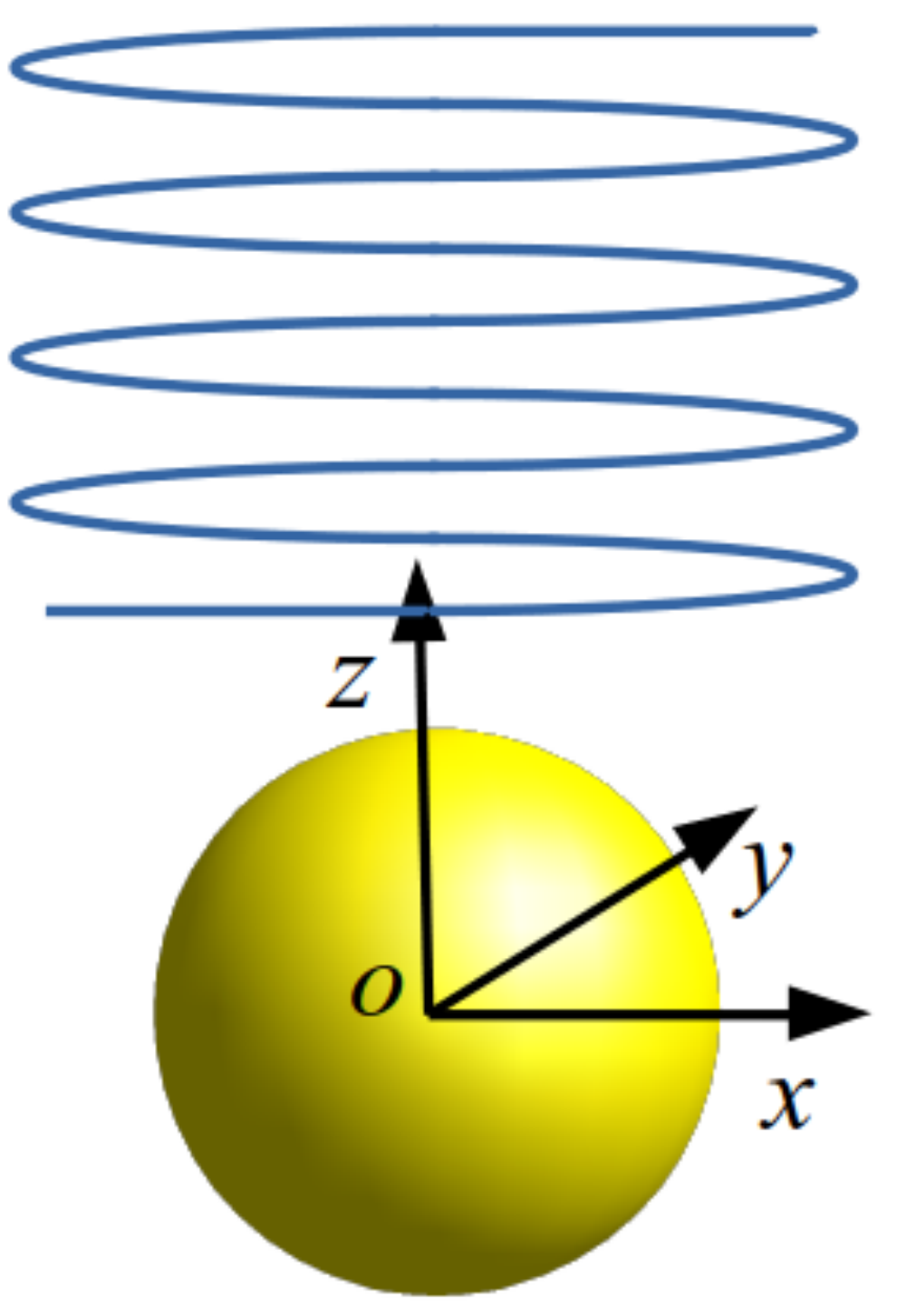}
	\caption{Plane electromagnetic wave incident on a conducting sphere.}{\label{physical_plot}}
\end{figure}
	Let us first understand the existence of symmetry in electromagnetic analysis with one example. 
	Consider one conducting sphere whose center is located at the origin as shown in Fig.~\ref{physical_plot}. 
	One electromagnetic plane wave, $\bE=E_xe^{-ikz}\hat\bi$, impinges on the sphere. 
	The geometry of the sphere is symmetric about the origin. 
	The incident wave has field component in $x$-direction and the wave is travelling from $z$-direction. 
	Thus, the incident wave is symmetric about the $xz$-plane. 
	There will be electromagnetic scattering by the conducting sphere. 
	As the geometry and the loading condition (incident wave) is symmetric about the $xz$-plane, the scattered electromagnetic field also will be symmetric about the $xz$-plane. 

When there exists some plane of symmetry, we can consider half of the domain on any one side of the symmetry plane for our numerical analysis. 
	In that case, one additional boundary surface gets created in the new half domain in form of the symmetric plane. 
	Proper boundary condition should be applied on that plane.
	
	Because of symmetry, always normal electric field to the symmetric plane should be zero. 
	This is the required boundary condition on the symmetric plane for electromagnetic analysis. 
	Thus, for the above example, on the symmetric plane ($xz$-plane), we have to satisfy $E_y = 0$. 
	In potential formulation, it is not very straight forward to satisfy $\bE\cdot \bn = 0$ as we have to eventually satisfy $(\bA\cdot \bn + \del\psi\cdot\bn) = 0$ .

	In this work, we have proposed a novel method of implementation of the symmetric boundary condition, $(\bA\cdot \bn + \del\psi\cdot\bn) = 0$. 
	Here, a thin patch of finite elements is attached to the symmetric plane, which extend our numerical domain. 
	The patch thickness $t$ should be very small as compared to the other dimensions of the domain. 
	For this extended domain, on the boundary surface which is parallel to the symmetric plane, $\bA\cdot \bn = 0$  is implemented explicitly on all the surface nodes. 
	In addition to that, for all the volume nodes inside this thin patch, $\psi = 0$ is applied explicitly. As the patch is very thin, it will result in $\del\psi\cdot\bn = 0$ on the symmetric plane.

\begin{figure}[pos=h!]
	\centering
	\includegraphics[width = 3.0 cm, height= 4.0 cm]{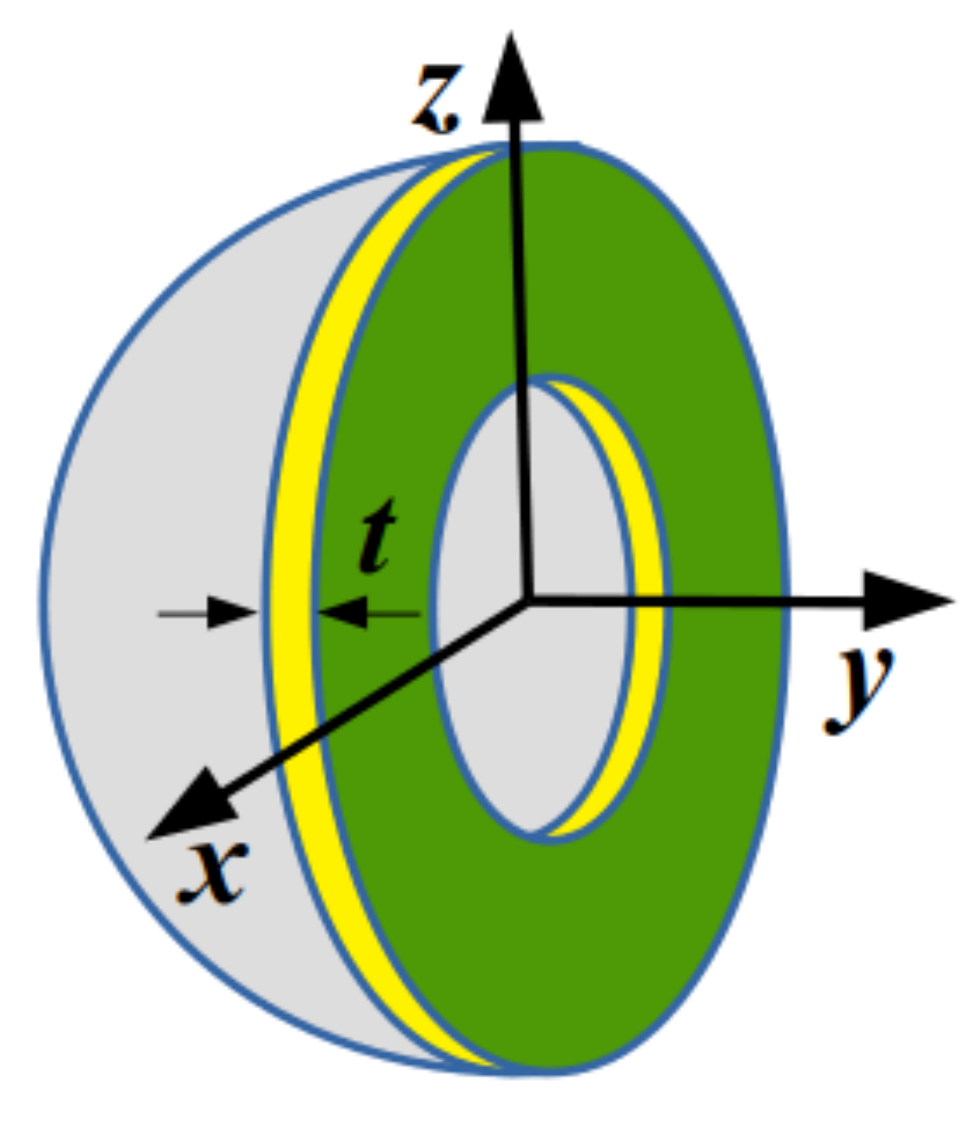}
	\caption{Sphere with a thin cylindrical patch of thickness, $t$.}{\label{spherepatch}}
\end{figure}
	
	In the context of our example, the full FEM analysis domain is one hollow sphere. 
	After considering the existing symmetry of the problem, we can take hollow hemisphere as our computational domain. 
	The flat face of the hollow hemisphere is the symmetric $xz-$plane. 
	As shown in Fig.~\ref{spherepatch}, one thin yellow patch of finite elements are attached to the symmetric flat face. 
	On the green surface which is parallel to the original symmetric plane, $A_y = 0$ is imposed for all the surface nodes. 
	For all the nodes inside the yellow volume of the cylindrical patch, $\psi = 0$ is applied so that we have $\del\psi\cdot\bn = 0$ on the symmetric face.
        
	In different examples, we have shown that the original converged solution field of the full domain is accurately predicted by the half domain with this novel implementation. 
	Detailed numerical analysis has been done, and electric fields are compared for both full and half domains along with the benchmark values. 
	In each and every case, results of both the domains are perfectly matching with the analytical/ benchmark values. 
	Moreover, it is also observed that in all the cases, with adoption of the present method, electric fields are simulated efficiently with less number of unknowns as compared to the full domain analysis. 
	Thus, the proposed method can be computationally efficient in electromagnetic analysis. 
	We have implemented the proposed method in both harmonic and transient electromagnetic analysis. 
	The proposed method has also been extended in amplitude formulation ~\cite{harmonic} for harmonic electromagnetic analysis.

\section{Numerical examples with electromagnetic harmonic analysis}
\label{numerical_examples_harmonic}
 In this section, we have presented the efficacy of the proposed method of implementing symmetric boundary condition, while solving various examples using potential formulation in harmonic analysis. 27 node hexahedral brick elements (B27) and 18 node wedge elements (W18) are used. The results obtained from the implementation of the symmetric boundary condition in the half domain, are compared with the full domain results, and the analytical benchmark results.

\subsection{Scattering from a conducting sphere} \label{subscat}

As the first example, we have solved scattering problem from one conducting sphere~\cite{harmonic}. Here, one electromagnetic plane wave, $\bE_\text{inc} =E_0e^{-ikz}\be_x$, impinges on a conducting sphere of radius $a$. To implement FEM, our computational domain (air surrounding the conducting sphere) is one hollow sphere which is truncated at radius $R_\infty$. On this truncated surface, we apply Sommerfeld absorbing boundary condition. Conducting boundary condition is imposed on the inner spherical surface at radius $a$. Our interest is to find the scattered electric field in the hollow spherical air domain having inner and outer radii of $a$ and $R_\infty$ respectively. From the discussion in section~\ref{method2}, we can understand that the problem is symmetric about the $xz$-plane. Therefore, we consider half of the hollow sphere on one side of the $xz$ plane as our computational domain, and we apply symmetric boundary conditions explicitly on the symmetry face $(xz)$. Considering the results in~\cite{harmonic}, we discretize the hemispherical domain with $16\times12\times12$ B27 and W18 elements. As discussed in section~\ref{method2}, the symmetric boundary condition is implemented with a thin cylindrical patch having whereas for the thin cylindrical patch we used $16\times12\times1$ ($r\times\theta\times t$) elements. In Table~\ref{table_equations}, mesh details along with the total number of equations are presented for both full domain (existing formulation) and half domain with the thin patch (current formulation).

\begin{table}[pos=h!]
	\caption{Mesh and equation details for scattering from a conducting sphere.} \label{table_equations}
	\begin{center}
		\begin{tabular}{p{0.3\linewidth}p{0.2\linewidth}p{0.2\linewidth}p{0.2\linewidth}}\toprule
			\multicolumn{1}{l}{}  & \multicolumn{1}{l}{Half domain with thin patch}& \multicolumn{1}{l}{Full domain~\cite{harmonic}}  \\ \midrule
			\multicolumn{1}{l}{Mesh size} &       $16\times12\times12$ +    &     $16\times12\times24$ \\
			\multicolumn{1}{l}{$(r\times\theta\times\phi)$} &$16\times24\times1$&      \\ \midrule
			Number of equations    &80125    & 106858 \\ \bottomrule
		\end{tabular}
	\end{center}
\end{table}

\begin{figure}[pos=H]
	\centering
	\subfloat[$E_x$ (real)]{\includegraphics[trim={0cm 0cm 0cm 0cm},clip,width=0.39\columnwidth]{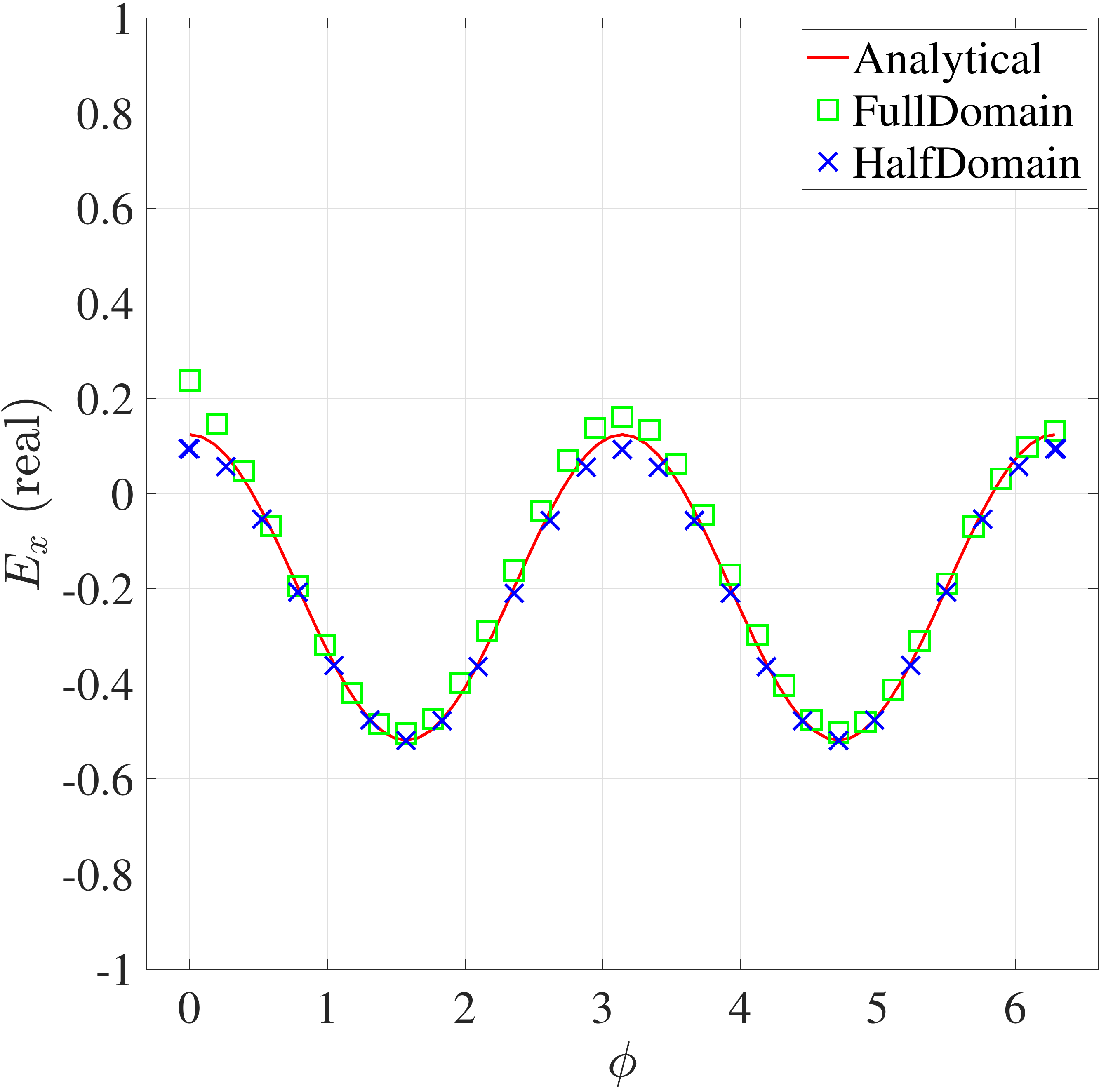}} \hspace{0.5cm}
	\subfloat[$E_x$ (imaginary)]{\includegraphics[trim={0cm 0cm 0cm 0cm},clip,width=0.39\columnwidth]{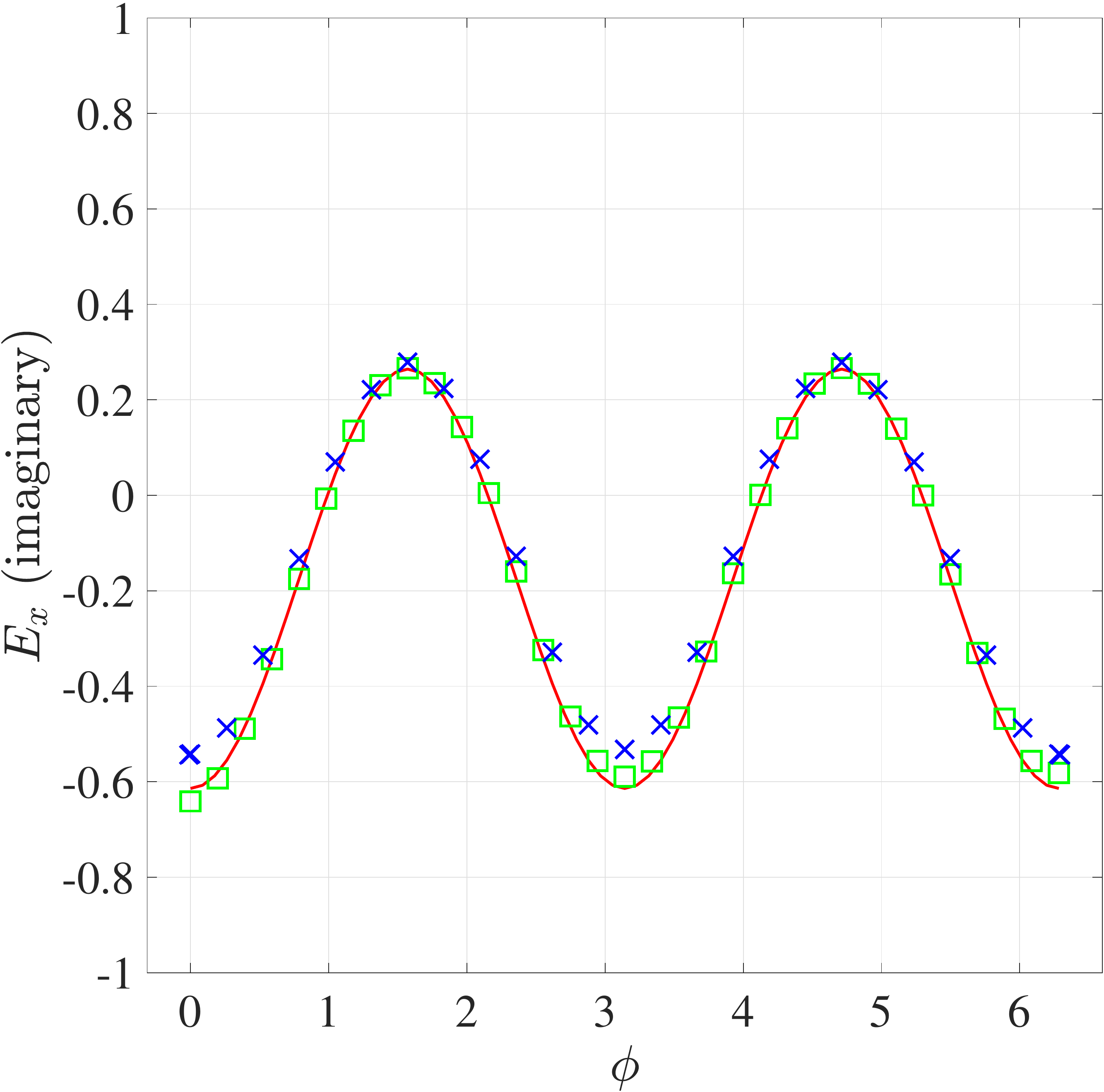}} \\
	\subfloat[$E_y$ (real)]{\includegraphics[trim={0cm 0cm 0cm 0cm},width=0.39\columnwidth]{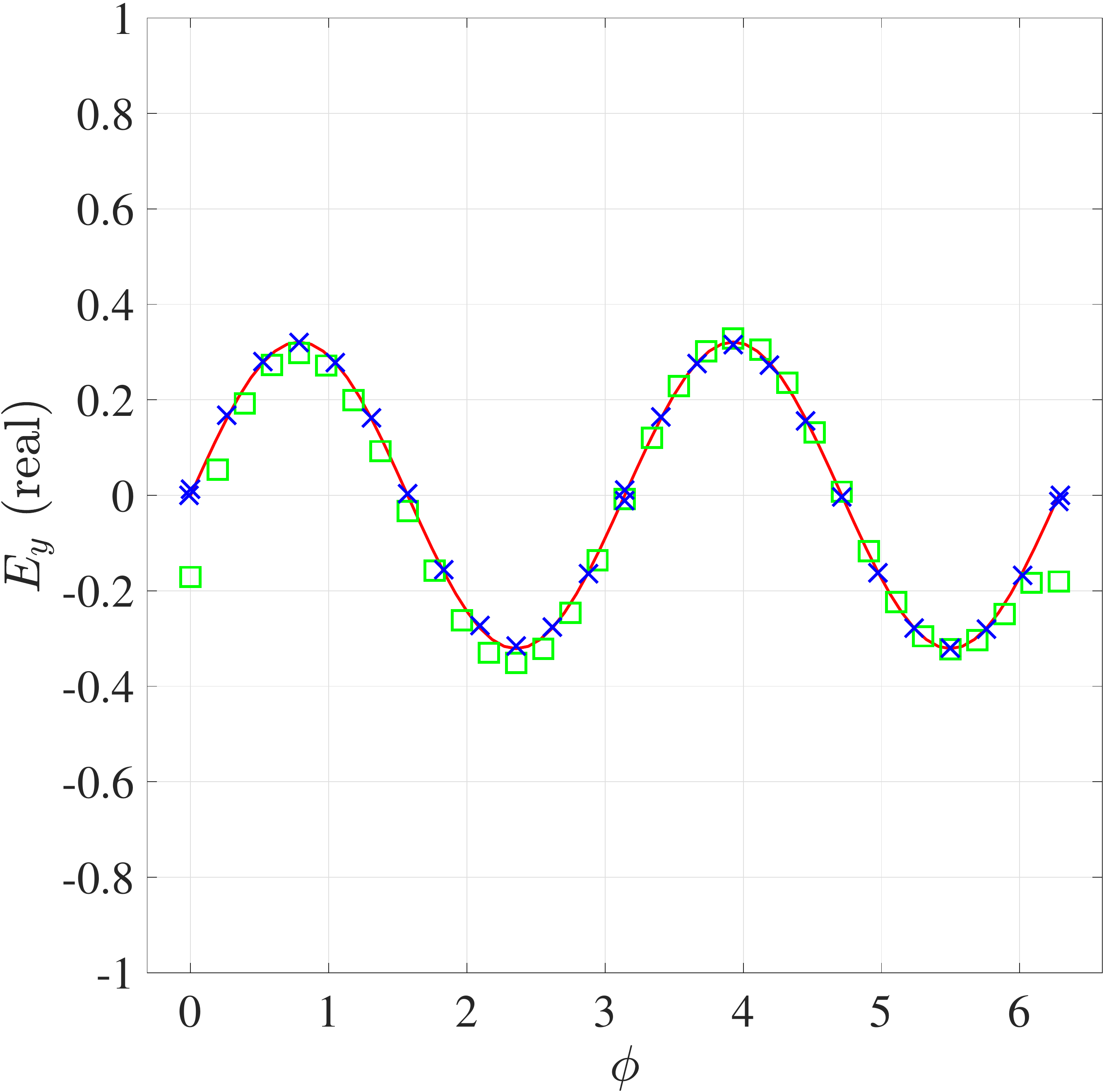}} \hspace{0.5cm} 
	\subfloat[$E_y$ (imaginary)]{\label{fig32b}\includegraphics[trim={0cm 0cm 0cm 0cm},width=0.39\columnwidth]{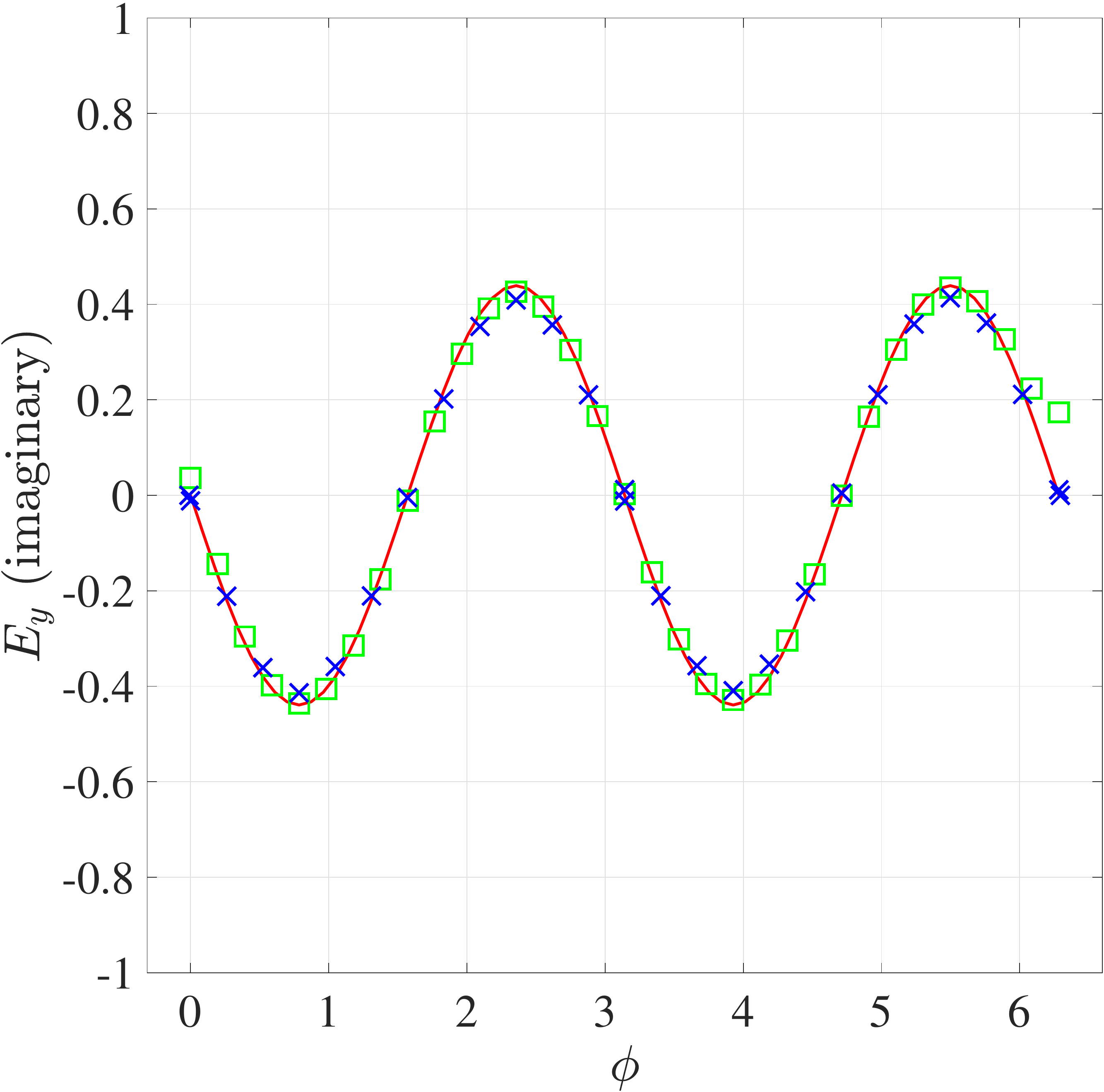}} \\
	\subfloat[$E_z$ (real)]{\includegraphics[trim={0cm 0cm 0cm 0cm},width=0.39\columnwidth]{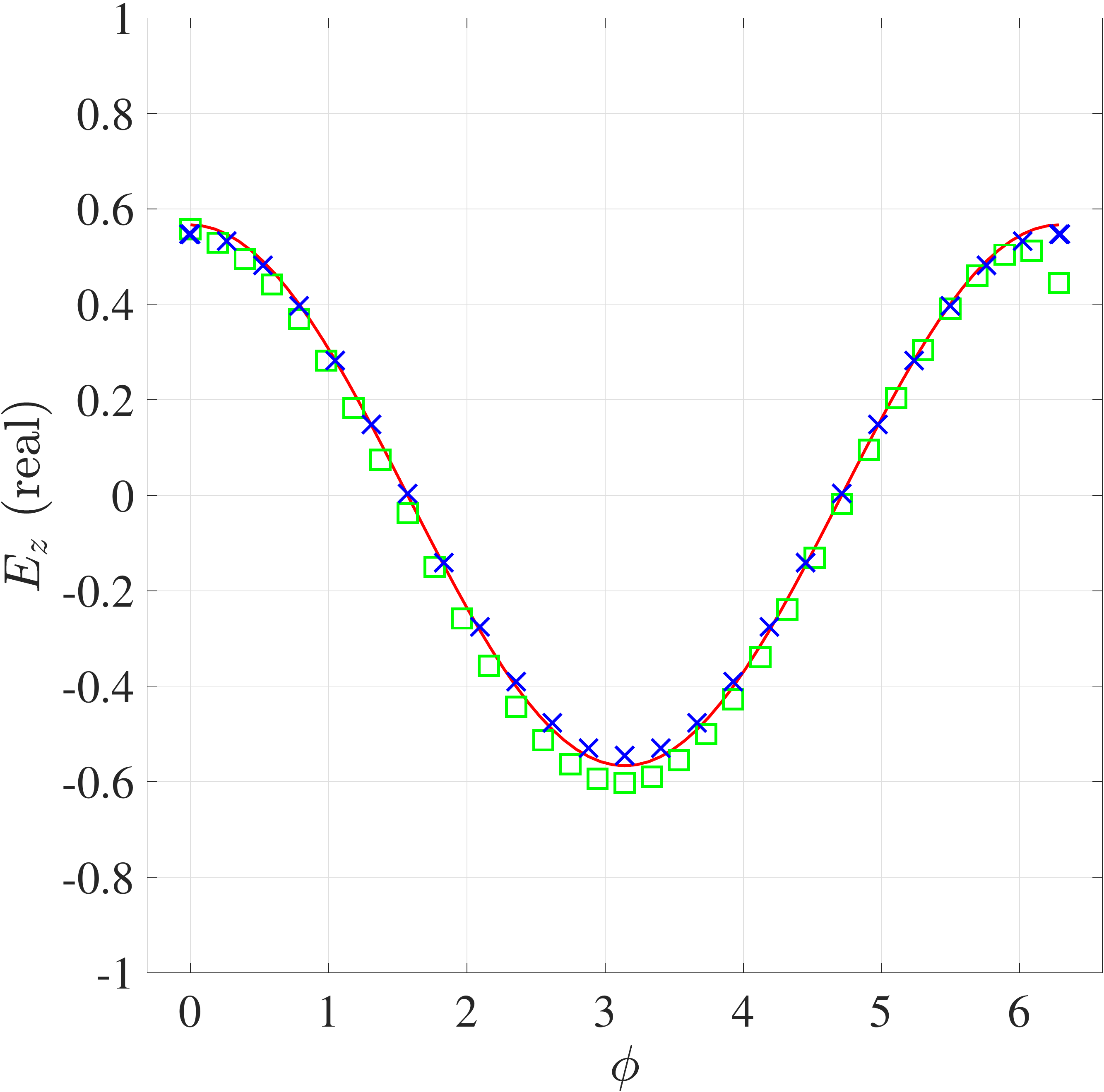}} \hspace{0.5cm}
	\subfloat[$E_z$ (imaginary)]{\includegraphics[trim={0cm 0cm 0cm 0cm},width=0.39\columnwidth]{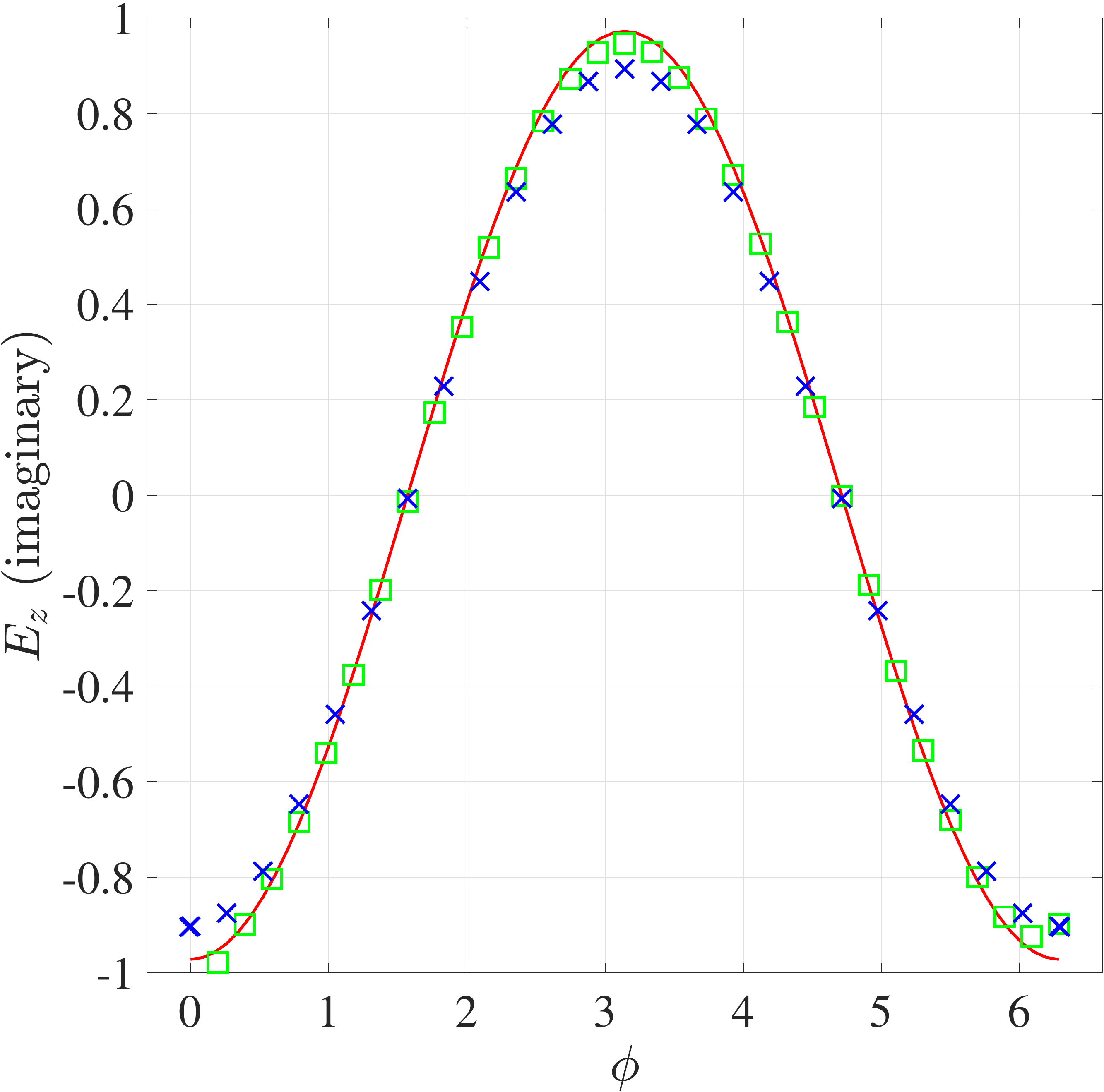}} \\
        \caption{Near field field variation along $\phi$ in the electric field for parameters $k_0r = 1.25$, $\theta = \pi/4$, $k_0a = 1$, $k_0R_\infty = 5$ for the scattering from a conducting sphere.}
	\label{fig_scat_sph_con_phi_near_ka1_Exp}
\end{figure}

\begin{figure}[pos=H]
	\centering
	\subfloat[$E_x$ (real)]{\label{fig41a}\includegraphics[trim={0cm 0cm 0cm 0cm},width=0.39\columnwidth]{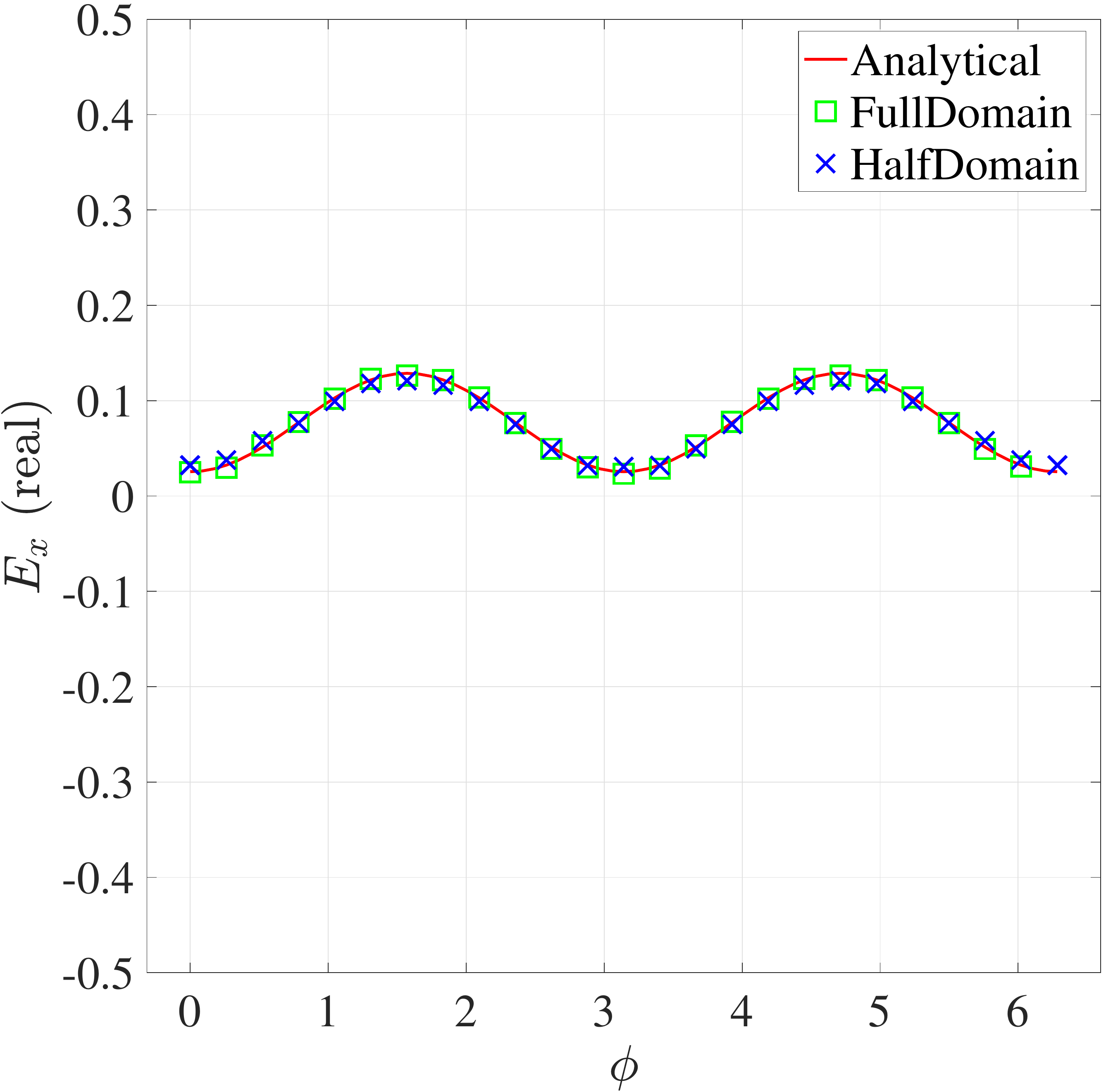}}\hspace{0.5cm} 
	\subfloat[$E_x$ (imaginary)]{\label{fig41b}\includegraphics[trim={0cm 0cm 0cm 0cm},width=0.39\columnwidth]{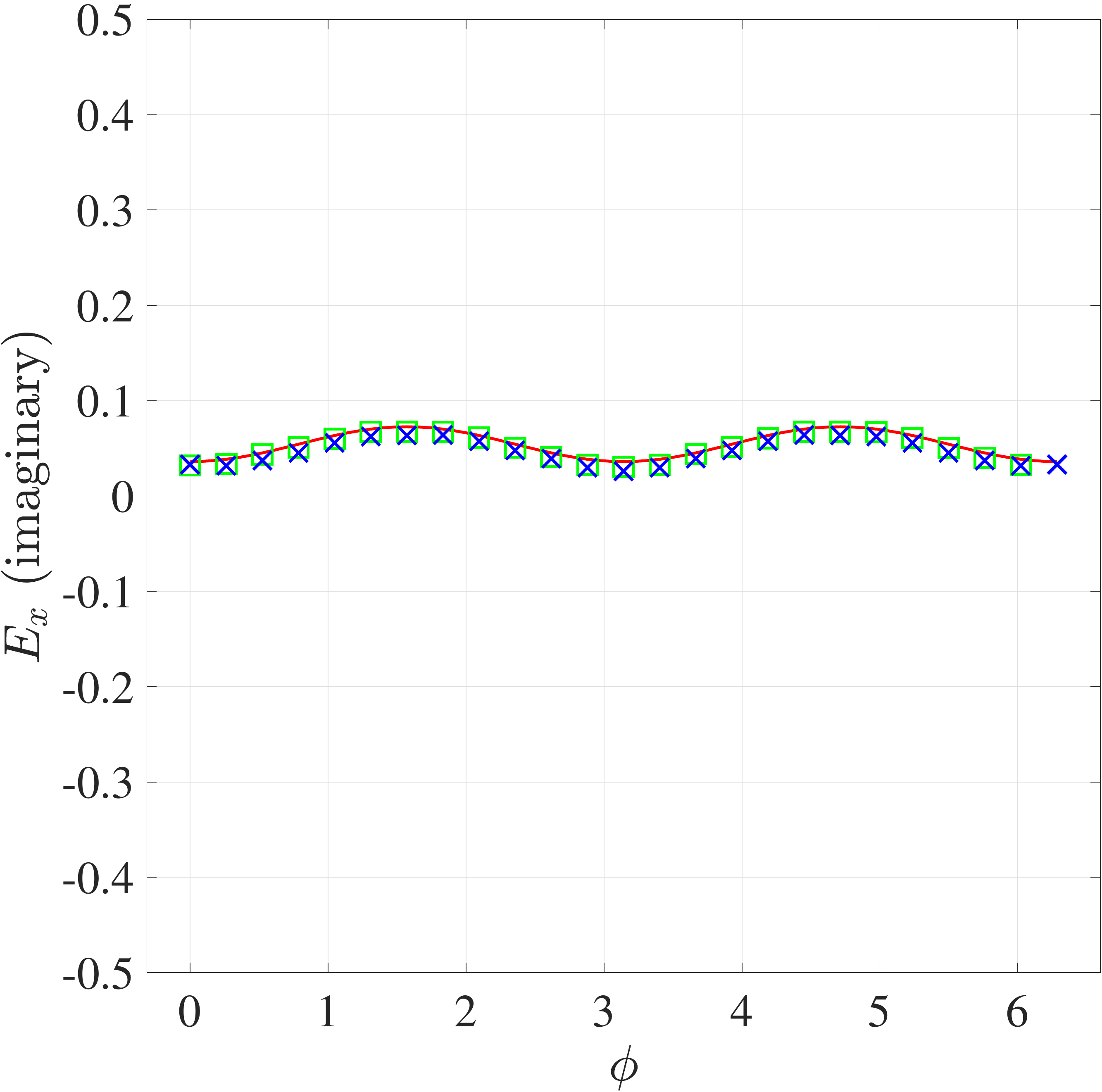}} \\
	\subfloat[$E_y$ (real)]{\label{fig42a}\includegraphics[trim={0cm 0cm 0cm 0cm},width=0.39\columnwidth]{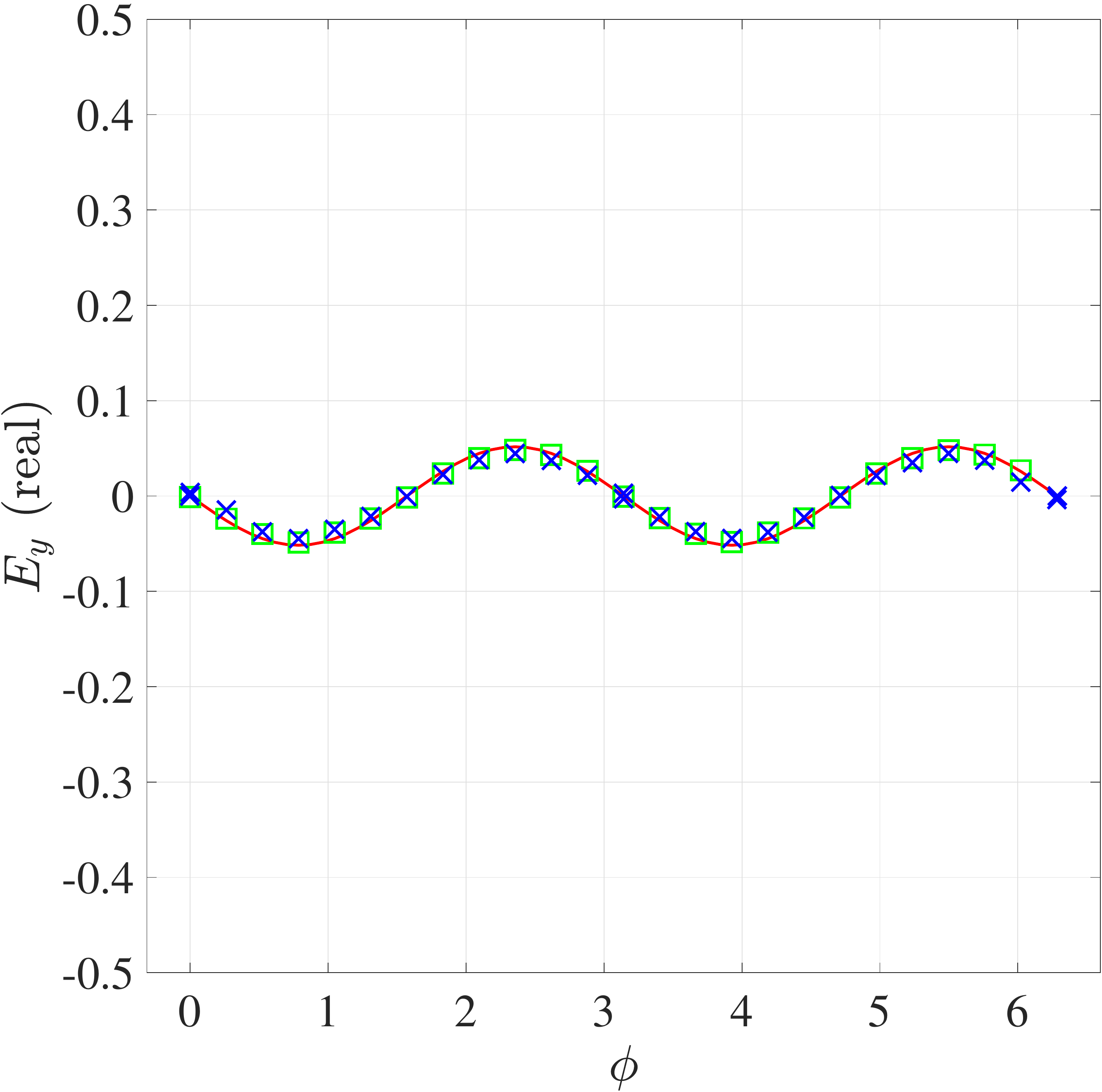}} \hspace{0.5cm}
	\subfloat[$E_y$ (imaginary)]{\label{fig42b}\includegraphics[trim={0cm 0cm 0cm 0cm},width=0.39\columnwidth]{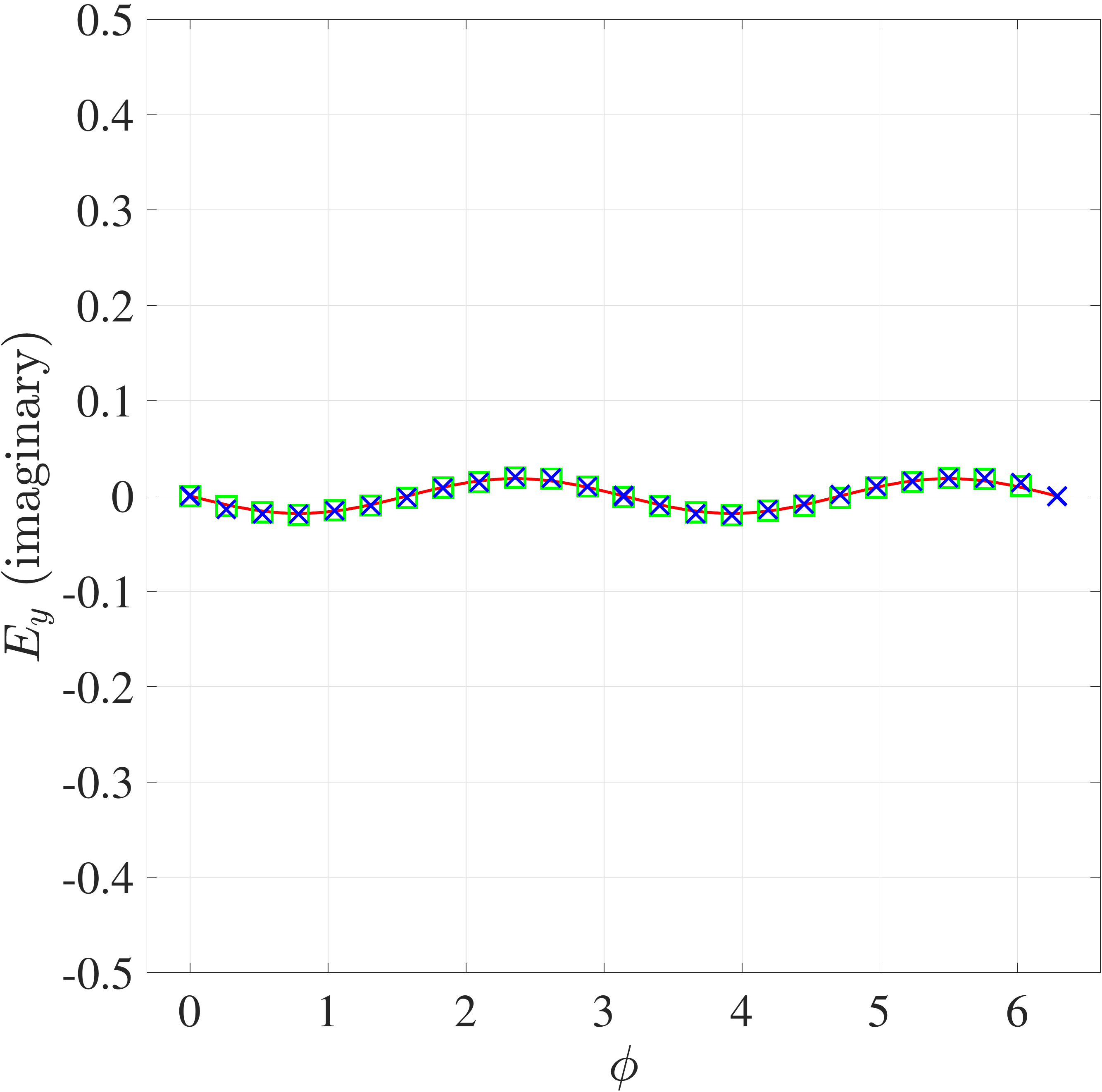}} \\
	\subfloat[$E_z$ (real)]{\label{fig43a}\includegraphics[trim={0cm 0cm 0cm 0cm},width=0.39\columnwidth]{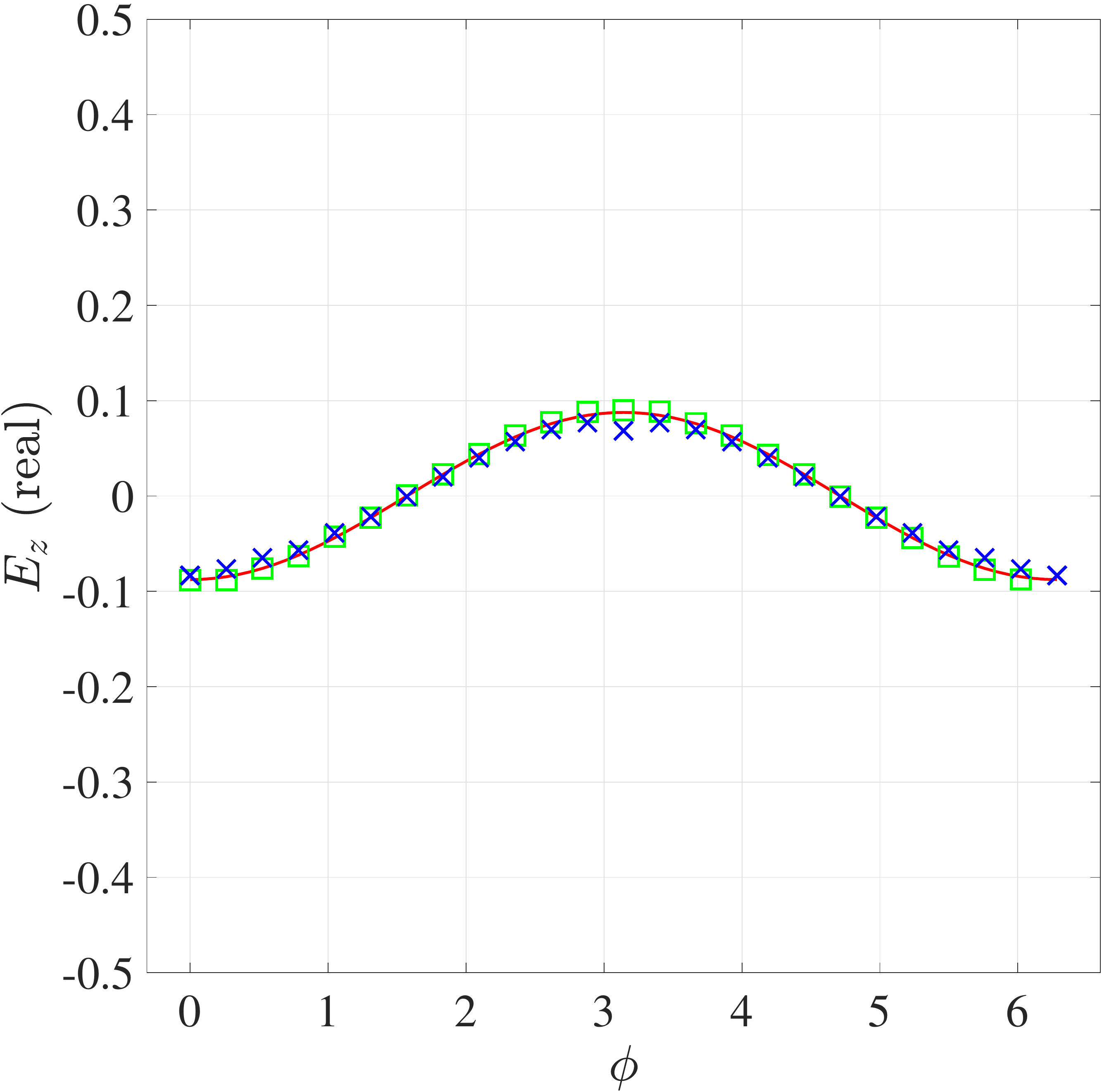}}\hspace{0.5cm}
	\subfloat[$E_z$ (imaginary)]{\label{fig43b}\includegraphics[trim={0cm 0cm 0cm 0cm},width=0.39\columnwidth]{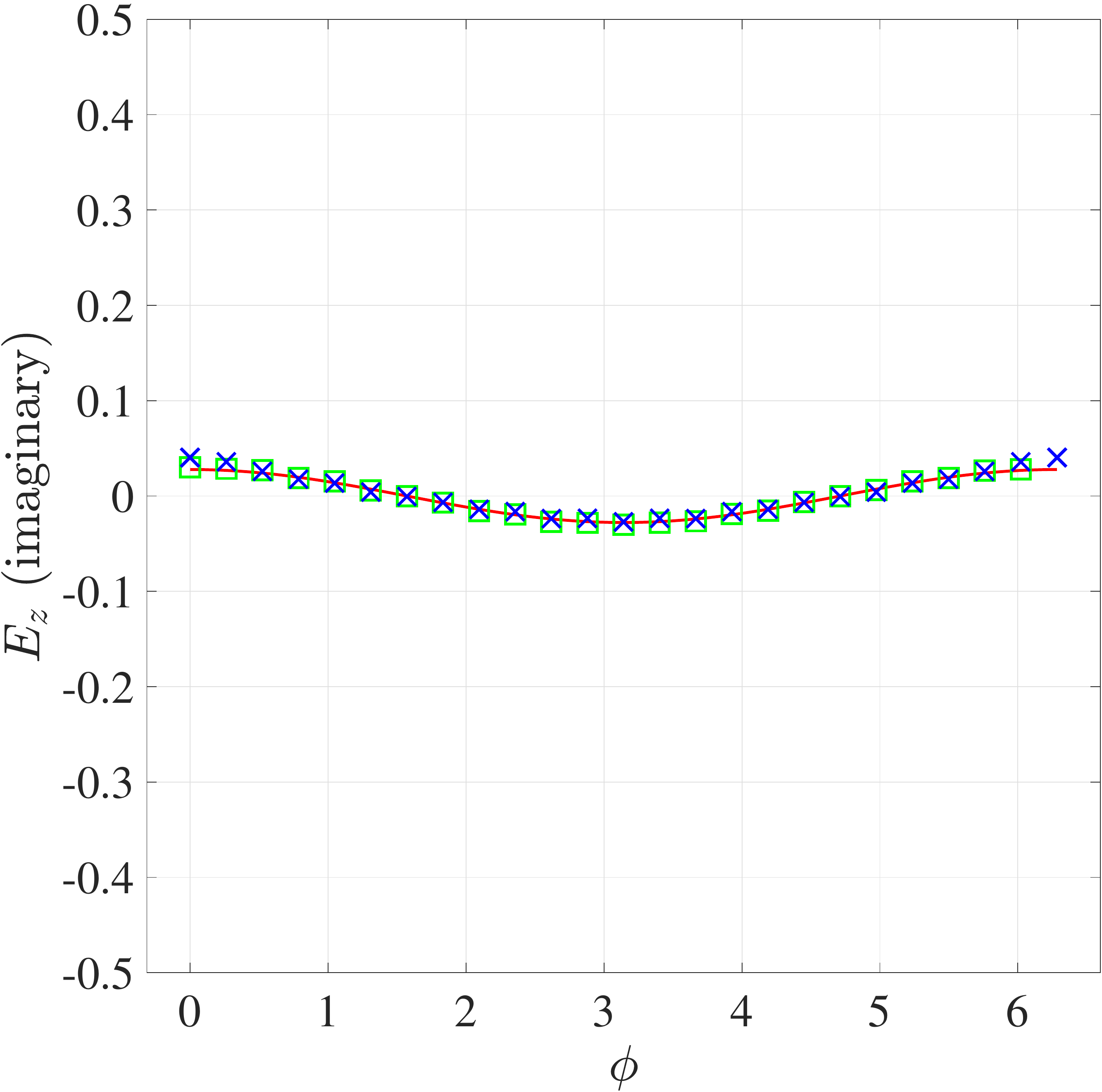}}\\
	\caption{Far field variation along $\phi$ in the electric field for parameters $k_0r = 4.75$, $\theta = \pi/4$, $k_0a = 1$, $k_0R_\infty = 5$ for the scattering from a conducting sphere.}
	\label{fig_scat_sph_con_phi_farr_ka1_Exp}
\end{figure}

We have considered $k_0a=1$, $k_0R_{\infty}=5$ with $E_0=1$. Fig.~\ref{fig_scat_sph_con_phi_near_ka1_Exp} and Fig.~\ref{fig_scat_sph_con_phi_farr_ka1_Exp} represent variations along $\phi$ for all three components of the scattered electric field at $\theta = \pi/4$, at near field ($k_0r$ = 1.15) and at far field ($k_0r$ = 4.75) respectively. During numerical analysis for the half domain, we directly solve for $0\le\phi\le\pi$. For the remaining half domain, we can use $\bE(\pi-\phi)=\bE(\phi)$ due to symmetry.  We can notice that half domain results are perfectly matching with the analytical benchmark results, while solving 25$\%$ less equations than the equivalent full domain.

\subsection{Scattering from a conducting ellipsoid}\label{conducntingellip}

This example is chosen to show that the proposed implementation of symmetric boundary condition also works for non spherical geometry. A wave of the form $e^{-ik_0z}\be_x $ impinges on one conducting ellipsoid having diameters 2$a$, 2$a$, and 2$c$ along $X$, $Y$ and $Z$ axes, respectively. The domain is truncated at a spherical surface given by $k_0R_\infty = 24$. We present our results for $k_0a = 12$ and $c = a/4$. In this problem, the symmetry boundary condition $\bE\cdot\bn=\bzero$ is implemented as described in section~\ref{method2}. The half annular ellipsoidal domain is meshed with $12\times20\times8$ ($r\times\theta\times\phi$) B27 and W18 elements and an additional cylindrical patch has $12\times40\times1$ ($r\times\theta\times t$) B27 elements. Then we have to solve 73,949 equations. Table~\ref{table_equations_ellipsoid} summarizes the mesh details of half domain along with the full domain.

\begin{table}[pos=h!]
	\caption{Mesh and equation details for scattering from a conducting ellipsoid.} \label{table_equations_ellipsoid}
	\begin{center}
		\begin{tabular}{p{0.3\linewidth}p{0.2\linewidth}p{0.2\linewidth}}\toprule
			\multicolumn{1}{l}{Mesh size}  & \multicolumn{1}{l}{half domain}& \multicolumn{1}{l}{full domain~\cite{harmonic}}  \\ \midrule
			\multicolumn{1}{l}{$(r\times\theta\times\phi)$} &     $12\times20\times8$ +    &     $12\times20\times16$ \\
			\multicolumn{1}{l}{$(r\times\theta\times t)$} &$12\times40\times1$&      \\ \midrule
			Number of equations &73949    & 123018  \\ \bottomrule
		\end{tabular}
	\end{center}
\end{table}

\begin{figure}[pos=H]
	\centering
	\subfloat[$E_x$ (real)]{\includegraphics[trim={0cm 0cm 0cm 0cm},width=0.39\columnwidth]{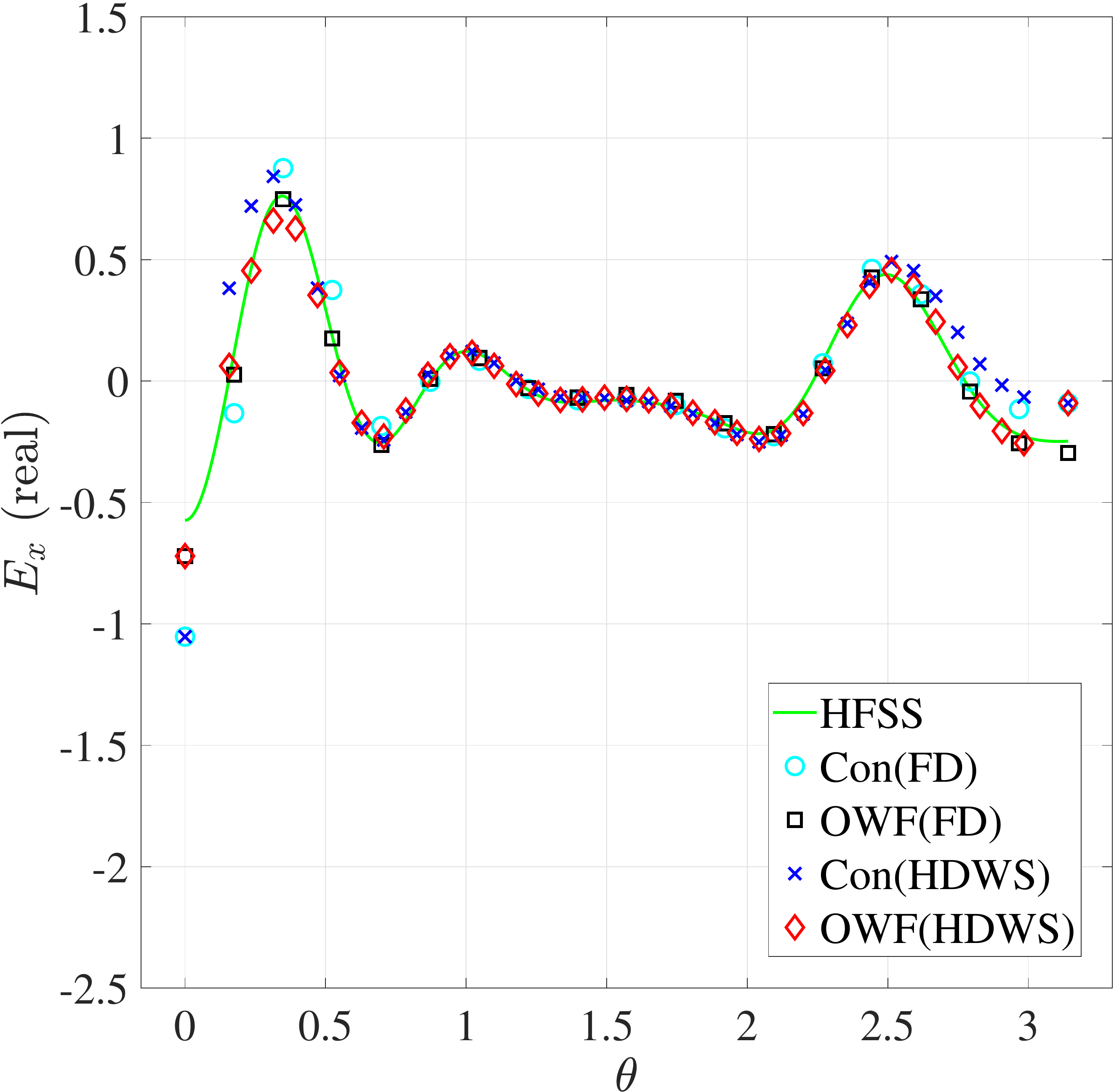}} \hspace{0.5cm} 
	\subfloat[$E_x$ (imaginary)]{\includegraphics[trim={0cm 0cm 0cm 0cm},width=0.39\columnwidth]{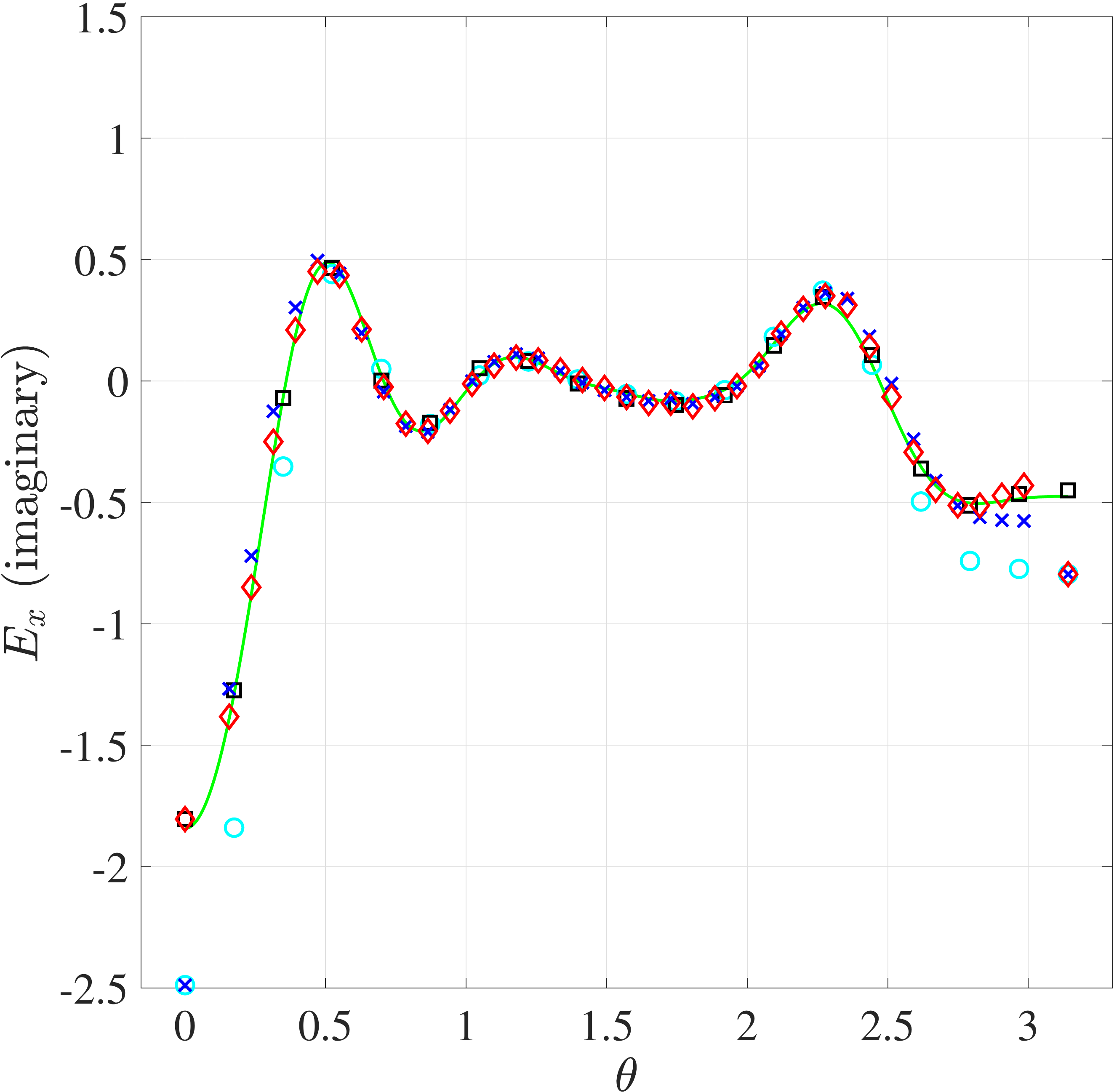}} \\
	\subfloat[$E_y$ (real)]{\includegraphics[trim={0cm 0cm 0cm 0cm},width=0.39\columnwidth]{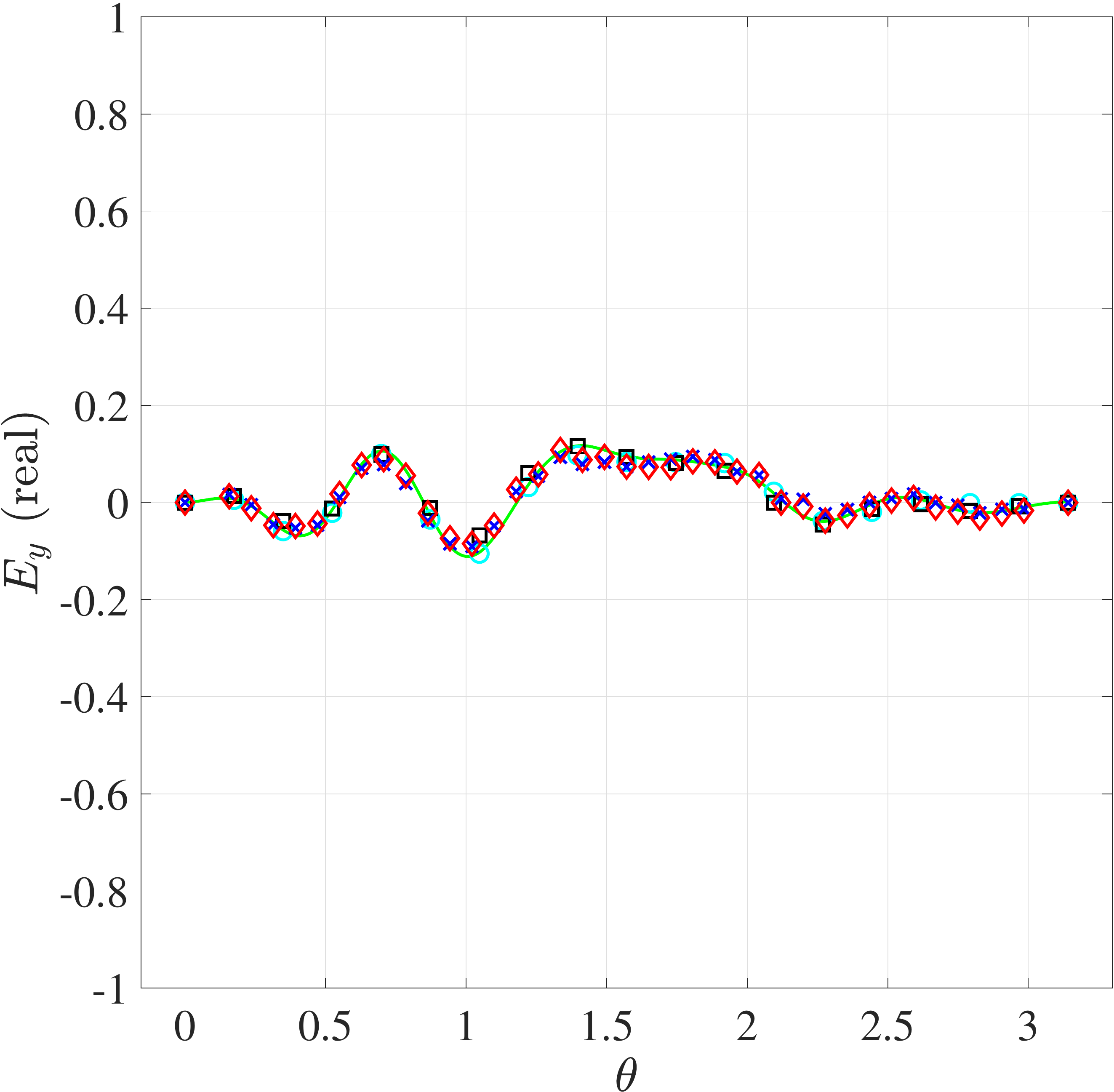}} \hspace{0.5cm}
	\subfloat[$E_y$ (imaginary)]{\label{fig32b}\includegraphics[trim={0cm 0cm 0cm 0cm},width=0.39\columnwidth]{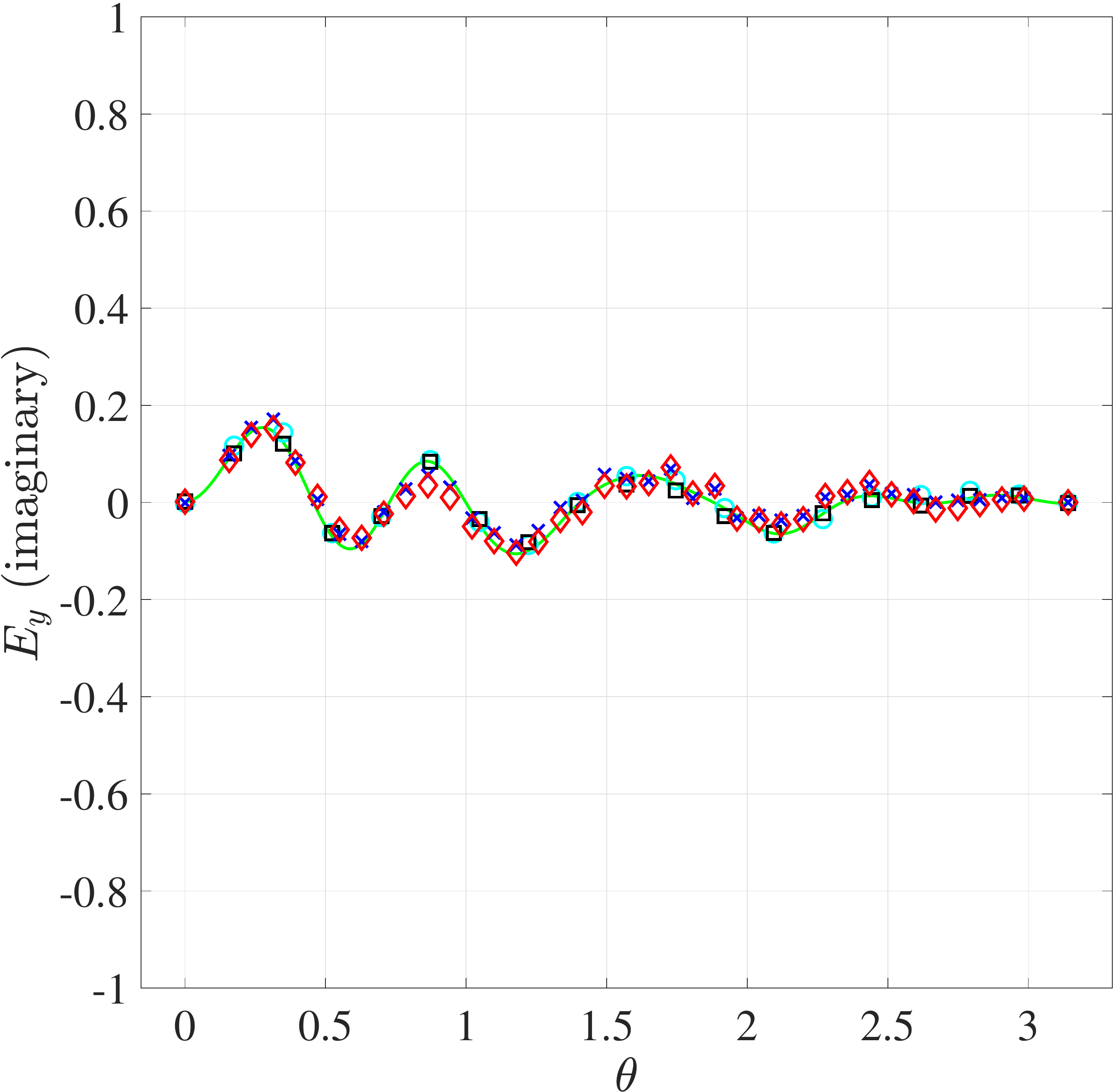}} \\
	\subfloat[$E_z$ (real)]{\includegraphics[trim={0cm 0cm 0cm 0cm},width=0.39\columnwidth]{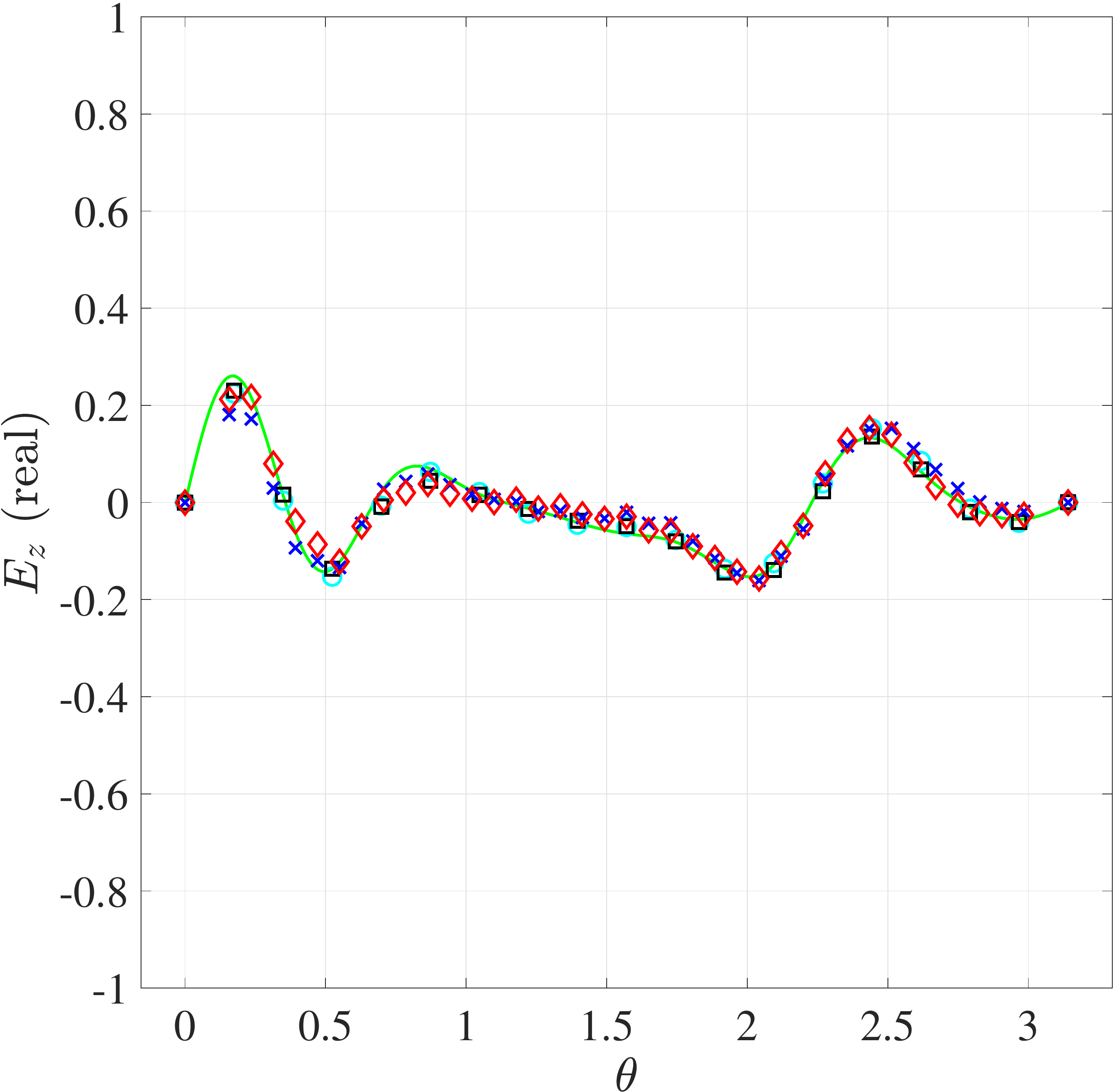}} \hspace{0.5cm}
	\subfloat[$E_z$ (imaginary)]{\includegraphics[trim={0cm 0cm 0cm 0cm},width=0.39\columnwidth]{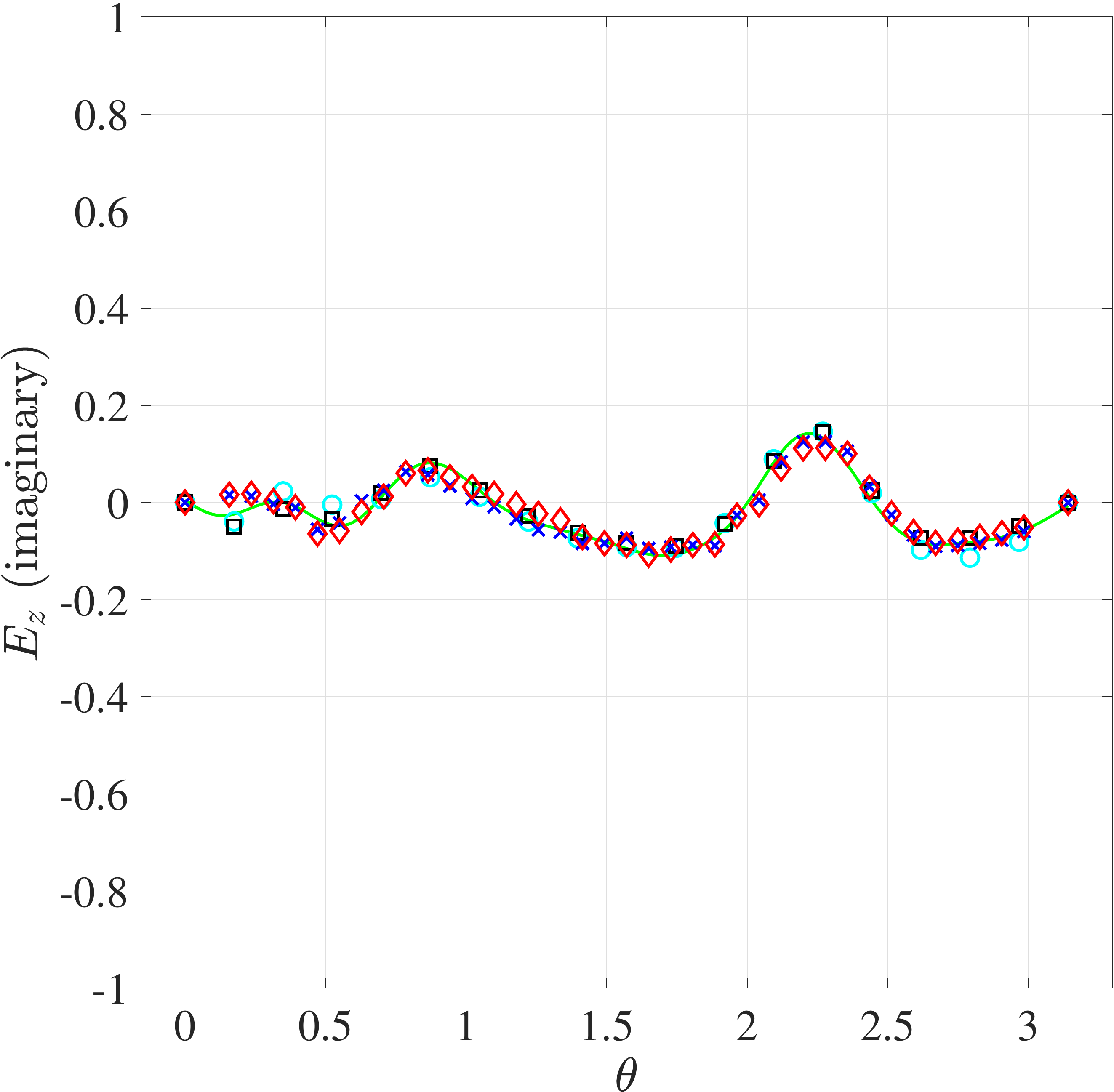}} \\
	\caption{Electric field variation along $\theta$ for $k_0r = 24$, $\phi = \pi/4$, $k_0a = 12$, $k_0R_\infty = 24$ for the scattering from a conducting ellipsoid.}
	\label{fig_scat_Elp_con_phi_far_ka3_Exp}
\end{figure}

We have taken $E_0 = 1, k_0a=12, k_0R_{\infty}=24$. The scattered electric field components are compared in Fig.~\ref{fig_scat_Elp_con_phi_far_ka3_Exp} along $\theta$ for $k_0r=24$ and $\phi=\frac{\pi}{4}$. We have compared the results of the half domain with symmetry (HDWS) in both conventional (Con) and amplitude formulation (OWF) with the converged HFSS~\cite{HFSS} result obtained using a very fine mesh (1,49,530 unknowns). In Fig.~\ref{fig_scat_Elp_con_phi_far_ka3_Exp}, we have also presented the respective full domain result for both Con and OWF formulations. In the current half domain analysis, we directly solve for $0\le\phi\le\pi$, and because of symmetry we have $\bE(2\pi-\phi) = \bE(\phi)$ for the remaining half domain. We obtain equivalent accuracy with half domain (thin patch) solving a system of equations having 40$\%$ less unknowns than that of full domain.

\subsection{Scattering from a dielectric sphere}\label{dielecsphere}

An incident wave of the form $\bE_{\text{inc}} = E_0e^{-ikz}\be_x$, impinges on a dielectric sphere. The relative permeability and relative permittivity of the dielectric sphere and the surrounding medium are denoted by $(\mu_1,\epsilon_1)$ and $(\mu_2,\epsilon_2)$, respectively. The analytical solution has been presented by Stratton~\cite{stratton2007electromagnetic}. In our numerical analysis, we assume the dielectric sphere to be in vacuum i.e., $\mu_2=\epsilon_2=1$ and it has material properties $\mu_1 = 1$ and $\epsilon_1 = 1.5$. We choose $ka = 8$, $kR_\infty = 24$, and $E_0=1$. For the proposed implementation of the half domain with symmetric boundary condition, 750 B27 and W18 elements are used to model the hemisphere whereas the thin cylindrical patch is modelled with 250 B27 elements. Mesh and equation details are given in Table~\ref{table_equations_dielectricsphere}.

\begin{table}[pos=h!]
	\caption{Mesh and equation details for scattering by dielectric sphere.} \label{table_equations_dielectricsphere}
	\begin{center}
		\begin{tabular}{p{0.3\linewidth}p{0.3\linewidth}p{0.2\linewidth}}\toprule
			\multicolumn{1}{l}{}  & \multicolumn{1}{l}{half domain}& \multicolumn{1}{l}{full domain~\cite{harmonic}}  \\ \midrule
			\multicolumn{1}{l}{Mesh details (no. of elements)} &     $750$(hemispherical domain) +    &     $1500$ \\
			\multicolumn{1}{l}{} &$250$(cylindrical patch)&      \\ \midrule
			Number of equations &28496    & 44692  \\ \bottomrule
		\end{tabular}
	\end{center}
\end{table}
\par
\begin{figure}[pos=H]
	\centering
	\subfloat[$k_0r = 6.4, \phi = \pi/6$ (real)]{\includegraphics[trim={0cm 0cm 0cm 0cm},width=0.39\columnwidth]{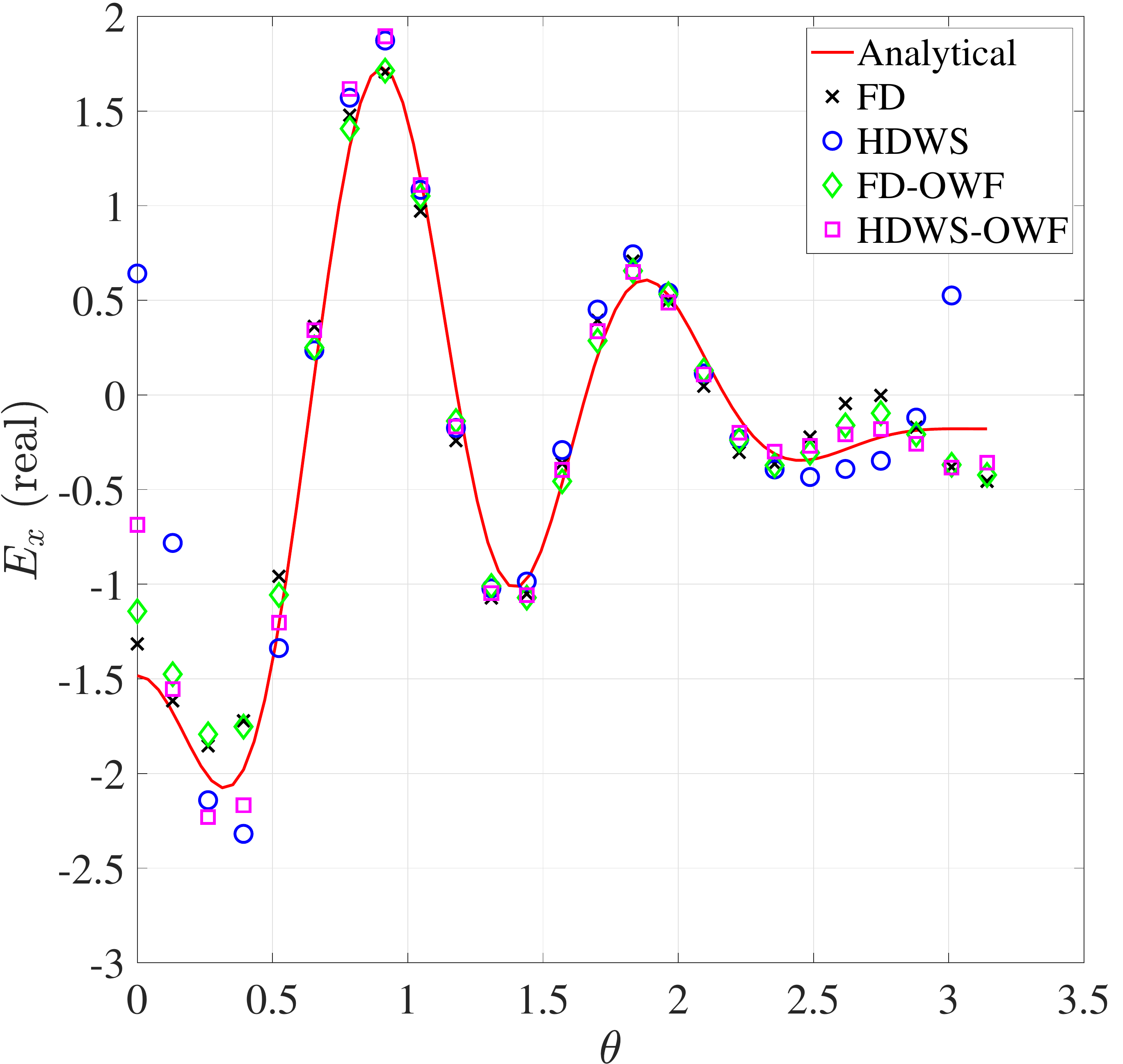}} \hspace{0.5cm}
	\subfloat[$k_0r = 6.4, \phi = \pi/6$ (imaginary)]{\includegraphics[trim={0cm 0cm 0cm 0cm},width=0.39\columnwidth]{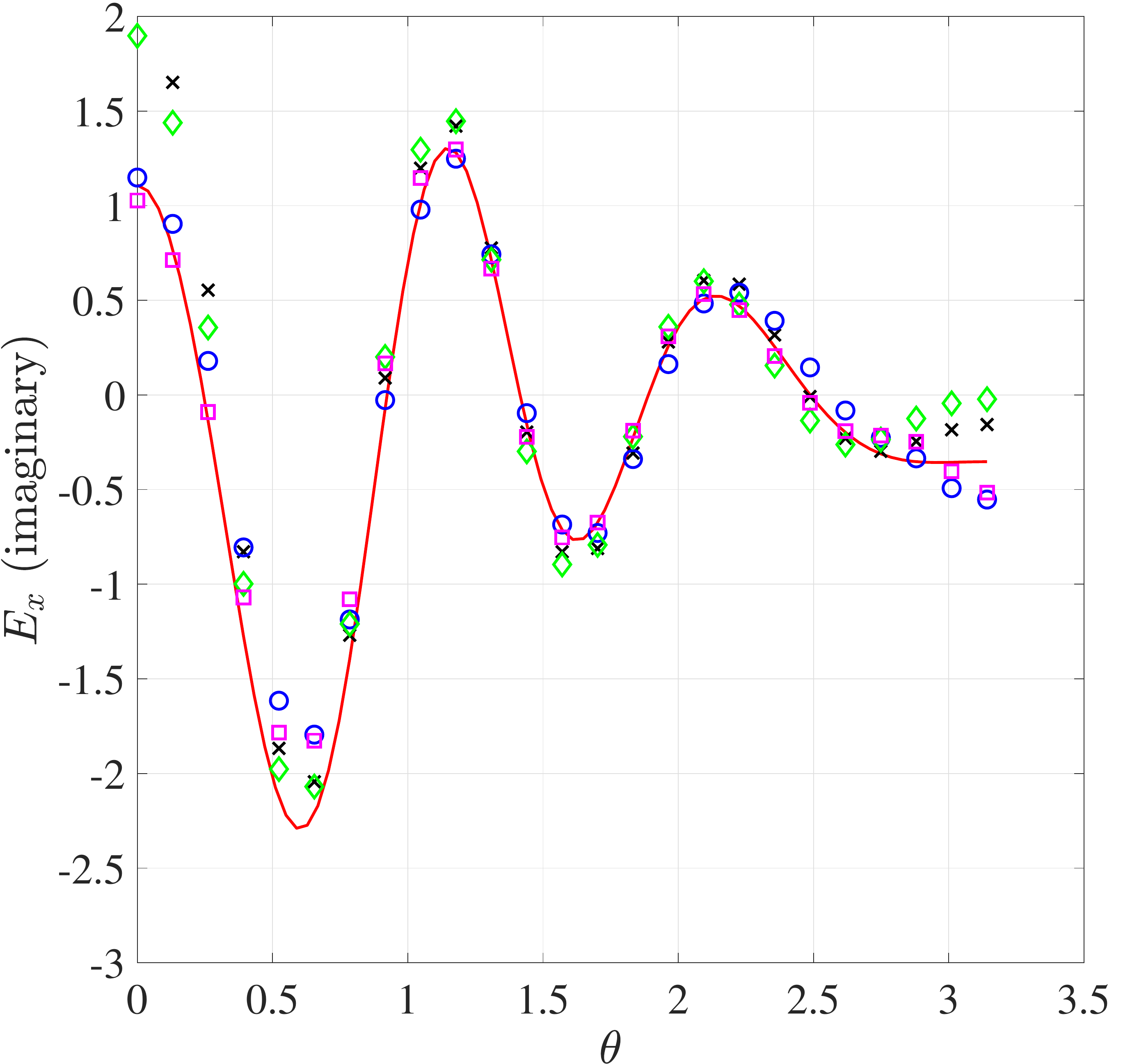}} \\
	\subfloat[$k_0r = 9.6, \phi = \pi/4$ (real)]{\includegraphics[trim={0cm 0cm 0cm 0cm},width=0.39\columnwidth]{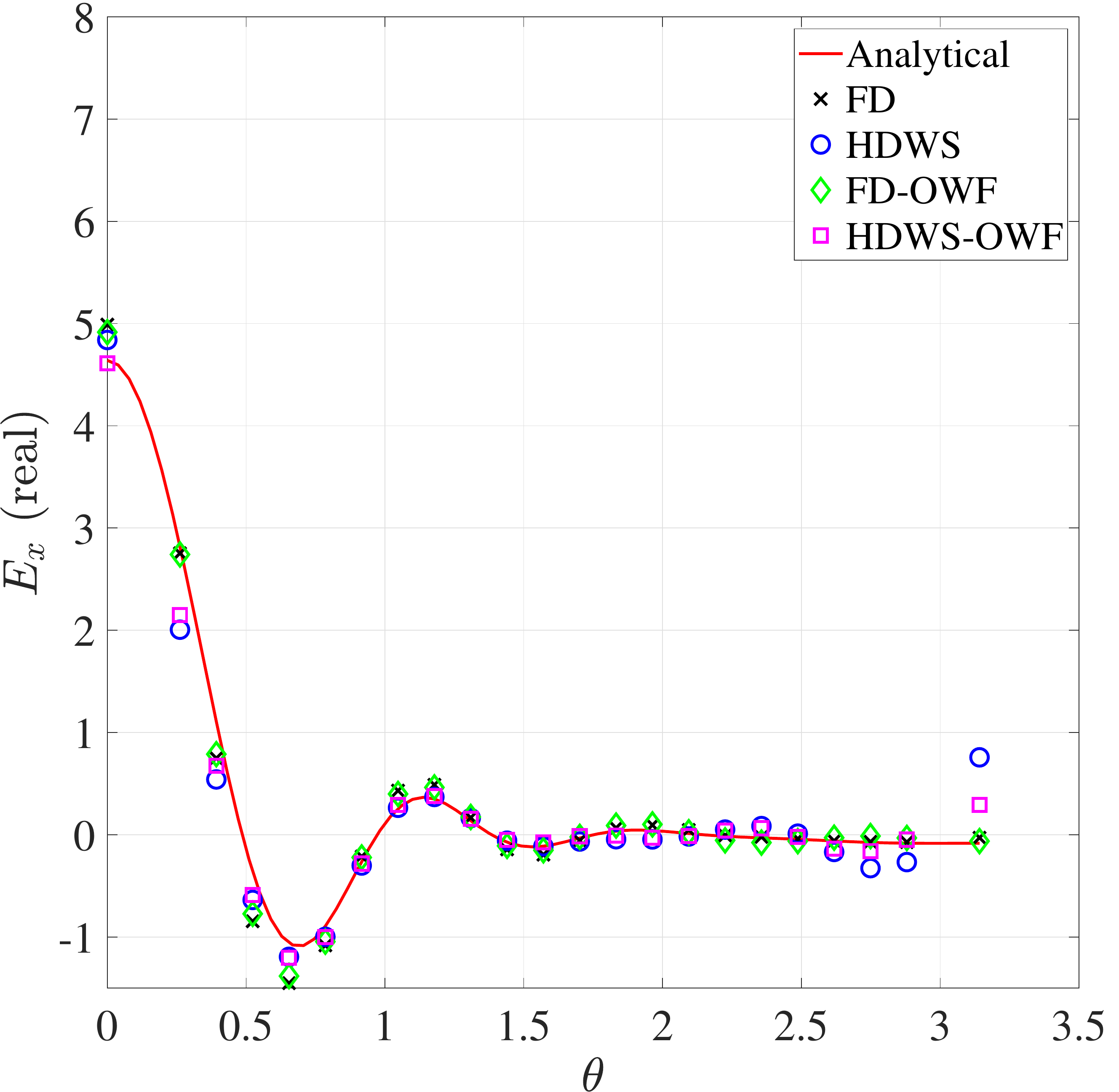}} \hspace{0.5cm}
	\subfloat[$k_0r = 9.6, \phi = \pi/4$ (imaginary)]{\includegraphics[trim={0cm 0cm 0cm 0cm},width=0.39\columnwidth]{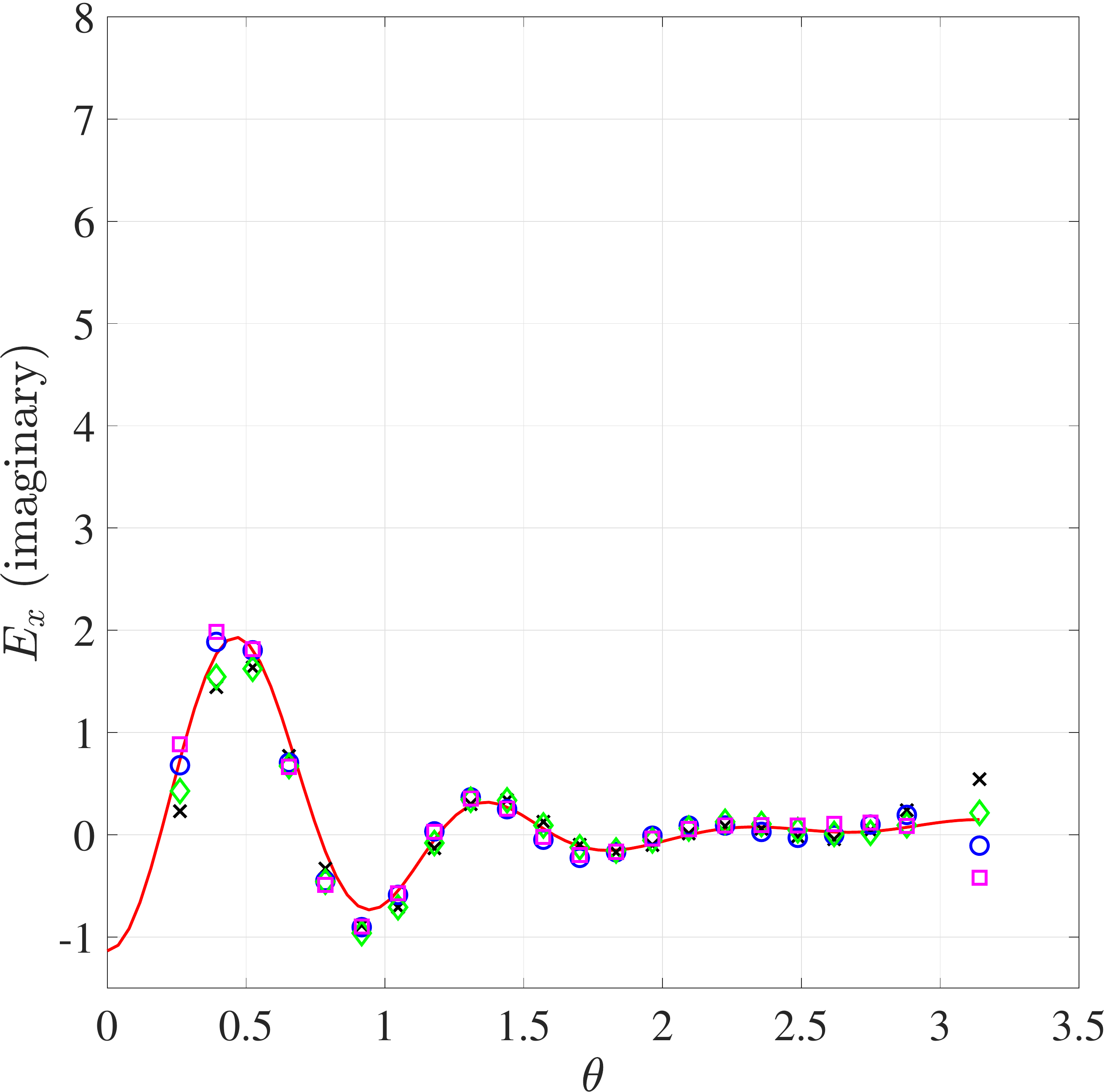}} \\
	\subfloat[$k_0r = 16, \phi = \pi/6$ (real)]{\includegraphics[trim={0cm 0cm 0cm 0cm},width=0.39\columnwidth]{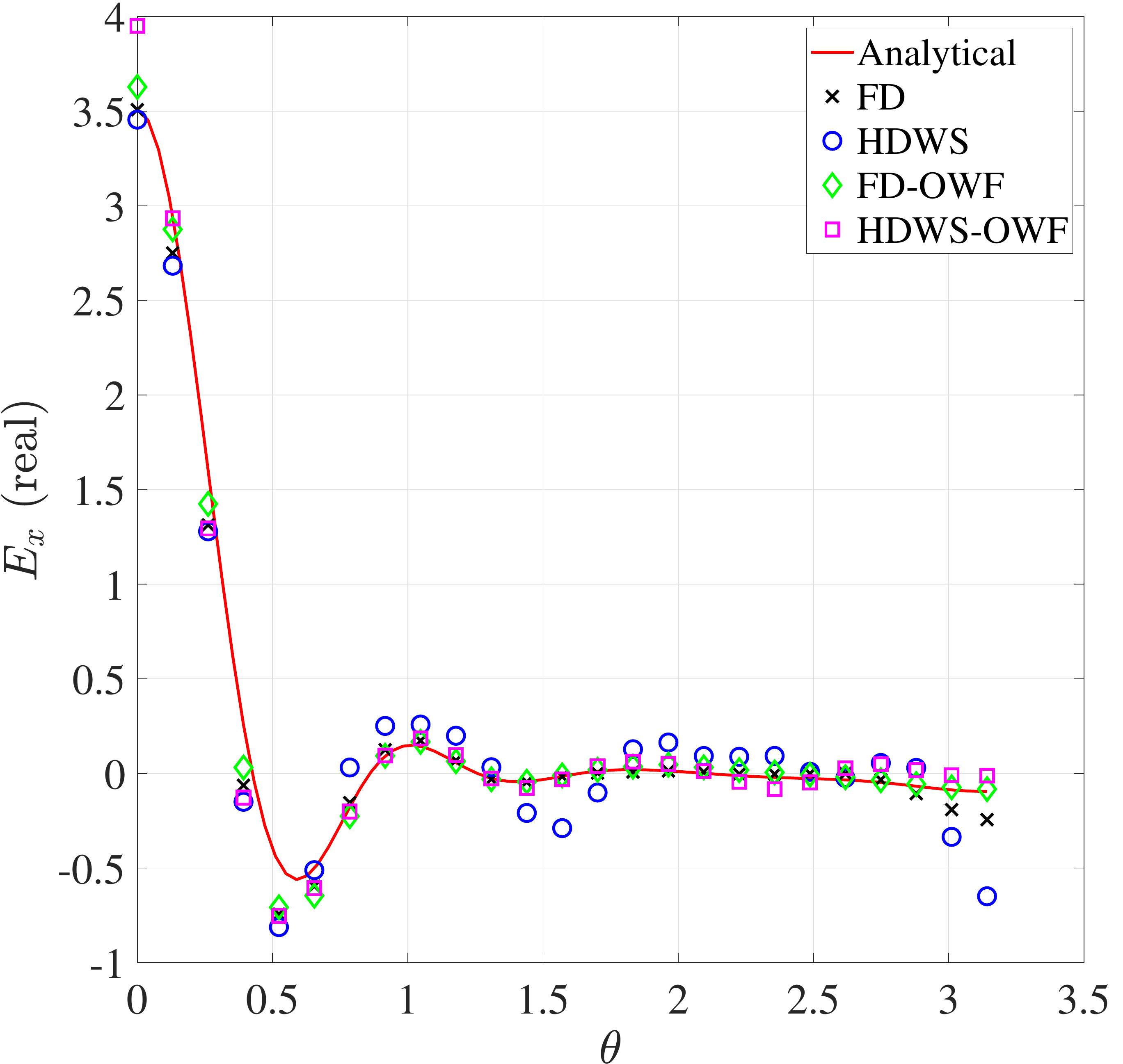}} \hspace{0.5cm}
	\subfloat[$k_0r = 16, \phi = \pi/6$ (imaginary)]{\includegraphics[trim={0cm 0cm 0cm 0cm},width=0.39\columnwidth]{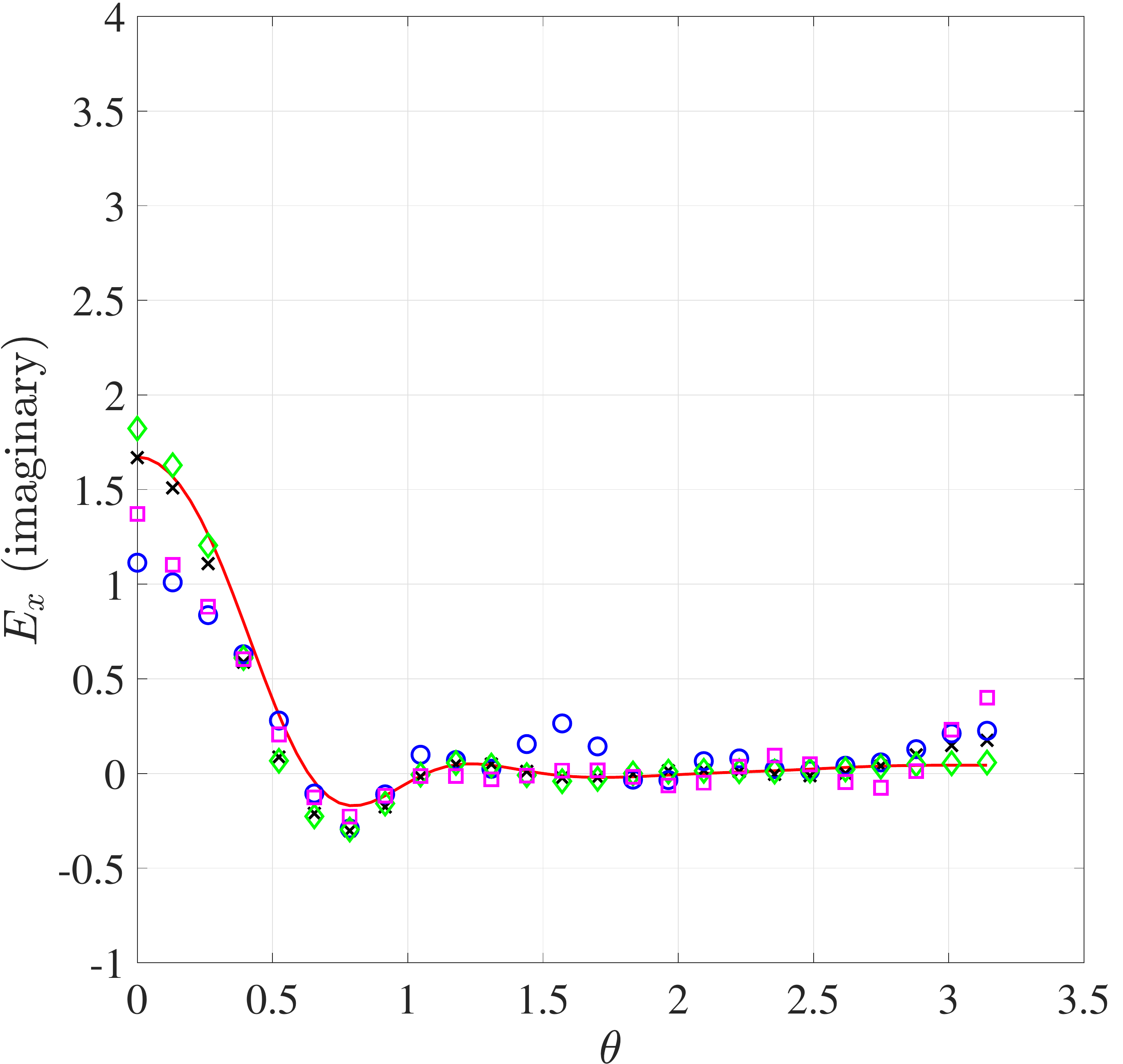}} \\
	\caption{Variation of scattered electric field $E_{x}$ along $\theta$ at different locations for the scattering from dielectric sphere problem.}
	\label{fig_scat_sph_dielec_th}
\end{figure}

 Fig.~\ref{fig_scat_sph_dielec_th} shows the results for the scattered electric field $E_x$ inside the dielectric sphere at $k_0r = 6.4$, at a point just outside the dielectric sphere ($k_0r = 9.6$), and at a point in the far-field ($k_0r = 16$) respectively. We have compared the results of both full domain (FD) and half domain with symmetry (HDWS) with analytical benchmark results as given by equation 34 \text{in}~\cite{harmonic}. In Fig.~\ref{fig_scat_sph_dielec_th}, results from both conventional and OWF formulations have been presented. In order to obtain same level of accuracy, we have to solve around 36$\%$ less no. of equations in HDWS as compared to the full domain, which clearly depict the computational efficacy of the proposed method. 

\section{Numerical examples with electromagnetic transient analysis}
\label{numerical_examples_transient}

In this section we apply the novel implementation of the symmetric boundary condition for various electromagnetic transient problems from~\cite{Nandy2018} to analyze the efficacy of the proposed method in time domain. These problems includes radiation inside a cube with conducting boundaries, scattering from a conducting and dielectric sphere. Again, we choose B27 and W18 elements to discretize the computational domains and we take $c=3\times10^8/\sqrt{\epsilon_r\mu_r} \text{ m/s}$.

\subsection{Electromagnetic radiation inside a cube with conducting walls} \label{cond_cube}
A cube of dimension $\pi\times\pi\times\pi$ with conducting walls is excited with $\bj$ and with initial $\bA$ values given \text{in}~\cite{Nandy2018} as
\begin{equation}\label{condbox}
\begin{split}
E_x &= 2\cos(x)\sin(y)\sin(z)\left[\cos(\omega t)-\sin(\omega t)\right], \\ 
E_y &= \sin(x)\cos(y)\sin(z)\left[\sin(\omega t)-\cos(\omega t)\right], \\ 
E_z &= \sin(x)\sin(y)\cos(z)\left[\sin(\omega t)-\cos(\omega t)\right], \\
H_x &= 0,\\
H_y &= \frac{-3}{\mu\omega}\cos(x)\sin(y)\cos(z)\left[\cos(\omega t)+\sin(\omega t)\right],\\
H_z &= \frac{3}{\mu\omega}\cos(x)\cos(y)\sin(z)\left[\cos(\omega t)+\sin(\omega t)\right],\\
A_x &= \frac{-2}{\omega}\cos(x)\sin(y)\sin(z)\left[\sin(\omega t)+\cos(\omega t)\right],\\
A_y &= \frac{1}{\omega}\sin(x)\cos(y)\sin(z)\left[\cos(\omega t)+\sin(\omega t)\right],\\
A_z &= \frac{1}{\omega}\sin(x)\sin(y)\cos(z)\left[\cos(\omega t)+\sin(\omega t)\right],\\
\psi &= 0,\\
j_x &= \left(\frac{2\epsilon\mu\omega^2-6}{\mu\omega}\right)\cos(x)\sin(y)\sin(z)\left[\cos(\omega t)+\sin(\omega t)\right], \\
j_y &= \left(\frac{3-\epsilon\mu\omega^2}{\mu\omega}\right)\sin(x)\cos(y)\sin(z)\left[\cos(\omega t)+\sin(\omega t)\right], \\
j_z &= \left(\frac{3-\epsilon\mu\omega^2}{\mu\omega}\right)\sin(x)\sin(y)\cos(z)\left[\cos(\omega t)+\sin(\omega t)\right].
\end{split}
\end{equation}

 We have taken $\omega=3\times10^8 \text{m/s}$, and simulated upto $4\times10^{-8} \text{sec}$ with a time step of $t_\Delta=1\times10^{-9} \text{sec}$.   

\begin{table}[pos=h!]
	\caption{Mesh and equation details for electromagnetic radiation inside a cube with conducting walls.} \label{table_equations_cond_cube}
	\begin{center}
		\begin{tabular}{p{0.3\linewidth}p{0.3\linewidth}p{0.2\linewidth}}\toprule
			\multicolumn{1}{l}{}  & \multicolumn{1}{l}{half domain}& \multicolumn{1}{l}{ful domain~\cite{Nandy2018}}  \\ \midrule
			\multicolumn{1}{l}{Mesh size } &  $2 \times 4 \times 2$ (cuboid) $+$    &     $8\times 8 \times 8$ \\
			\multicolumn{1}{l}{$(x \times y \times z)$} &$(1 \times 4 \times 2) + (2 \times 4 \times 1) + (1 \times 4 \times 1)$ (patches)&      \\ \midrule
			\multicolumn{1}{l}{Number of elements}& 36 & 64 \\ \midrule
			Number of equations & 890    & 1666  \\ \bottomrule
		\end{tabular}
	\end{center}
\end{table}

To implement the symmetric boundary conditions, we have taken one quarter of the original cube with $\frac{\pi}{2}\times\pi\times\frac{\pi}{2}$. On the symmetry faces i.e., on $xy$ and $yz$-planes, we have applied $\bE\cdot\bn=\bzero$ as explained in section~\ref{method2}, and on the remaining faces of the cuboid, conducting boundary conditions $\bE\bcross\bn=\bzero$ are applied. For transient analysis, 36 B27 elements are used to mesh the quarter and the thin patches, whereas full domain is meshed with 64 elements. Table~\ref{table_equations_cond_cube} mesh details for both the domains. In Fig.~\ref{cond_cube}, we have compared the electric field components at ($x =$ 1.1780, $y =$ 0.3926, $z =$ 0.7853), obtained from the proposed formulation with the existing full domain results and the analytical solution~\cite{Nandy2018}. Again, it is evident, by adopting the symmetric boundary conditions, 890 equations are sufficient for the perfect match with analytical benchmark as compared to 1666 unknowns for the full domain.

\begin{figure}[pos=h!]
 \centering
	\subfloat[$E_x$]{\includegraphics[trim={12cm 0cm 12cm 0cm},width=0.42\columnwidth]{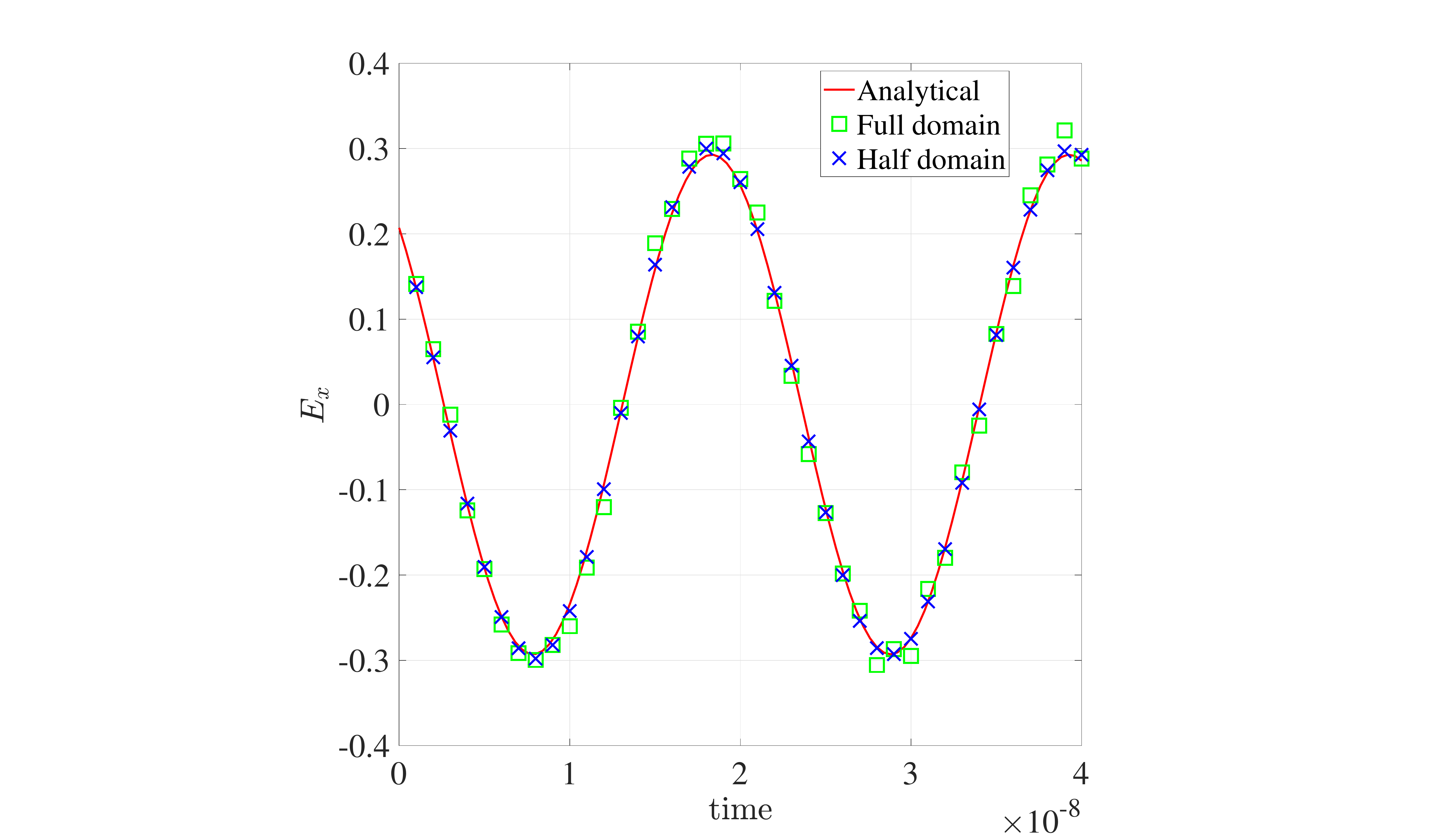}} \hspace{0.5cm}
	\subfloat[$E_y$]{\includegraphics[trim={0cm 0cm 0cm 0cm},width=0.45\columnwidth]{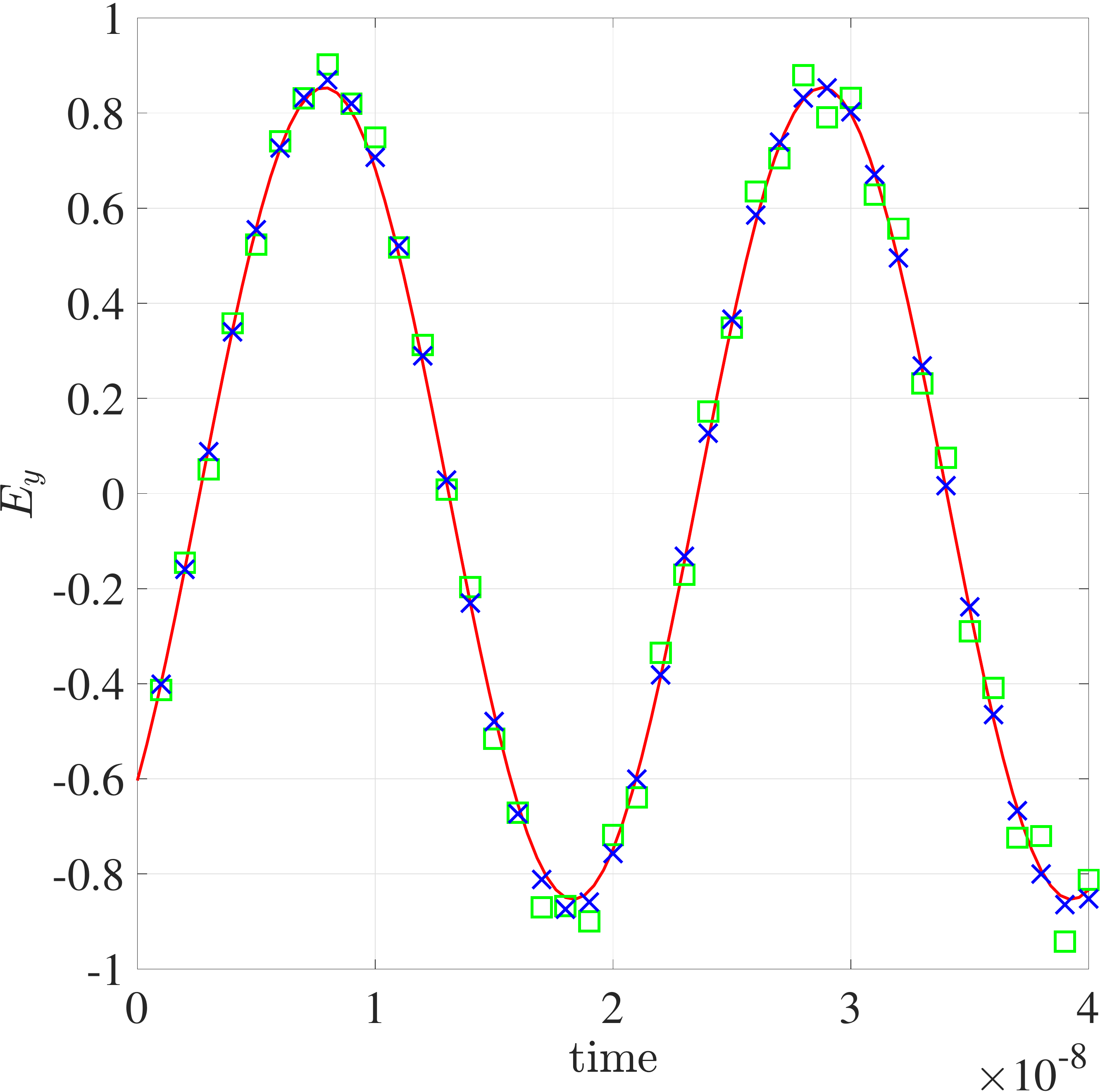}} \\
	\subfloat[$E_z$]{\includegraphics[trim={0cm 0cm 0cm 0cm},width=0.45\columnwidth]{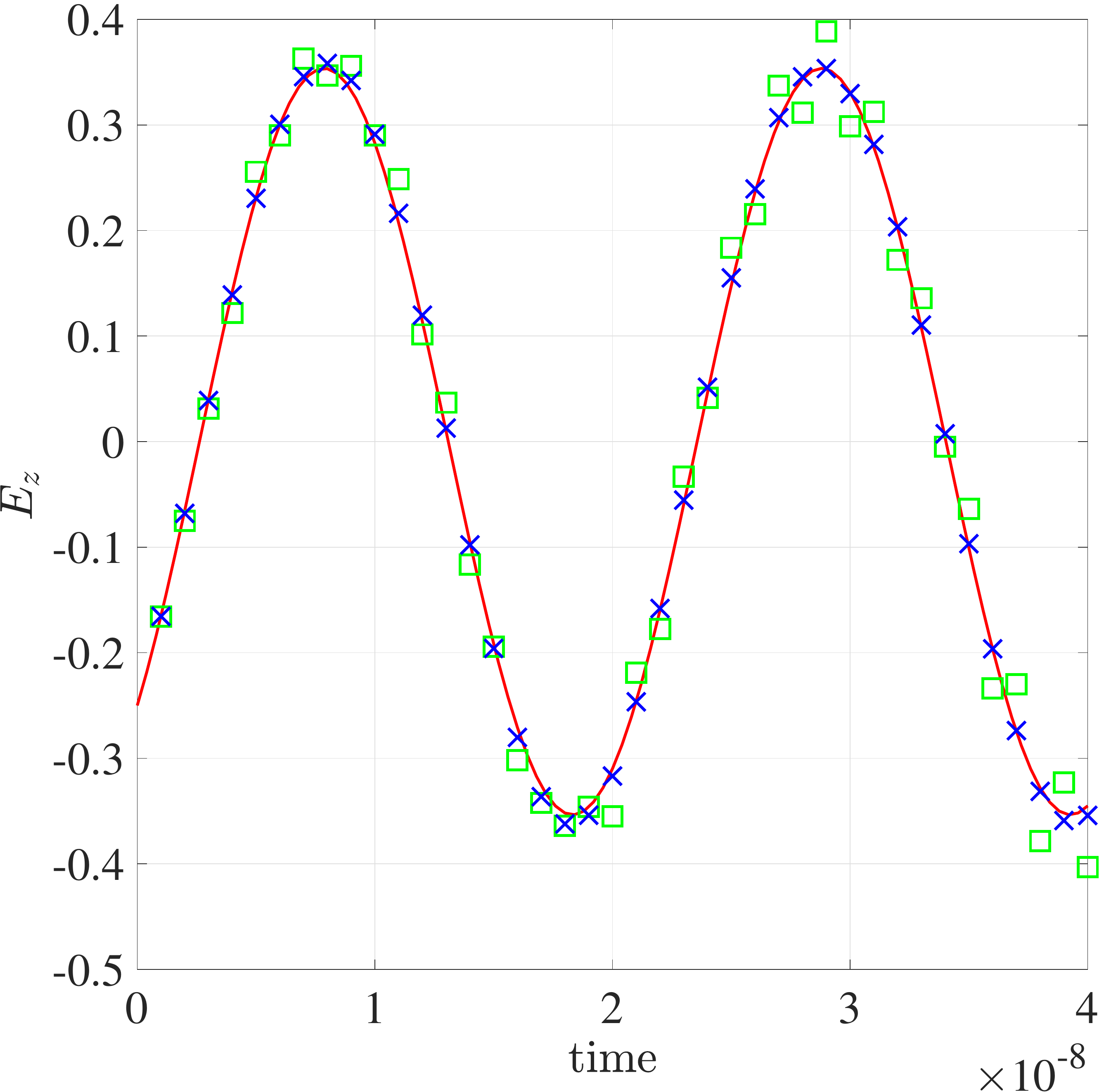}} \\ 
	\caption{Variation of the electric fields at $(x,y,z)=(1.1780,0.3926,0.7853)$ with respect to time for the cube with conducting walls.}
\label{cond_cube}
\end{figure}

\subsection{Electromagnetic scattering from a conducting sphere} \label{scat_cond_sphere}
\begin{figure}[pos=h!]
 \centering
	\includegraphics[trim={0cm 0cm 0cm 0cm},width=0.55\columnwidth]{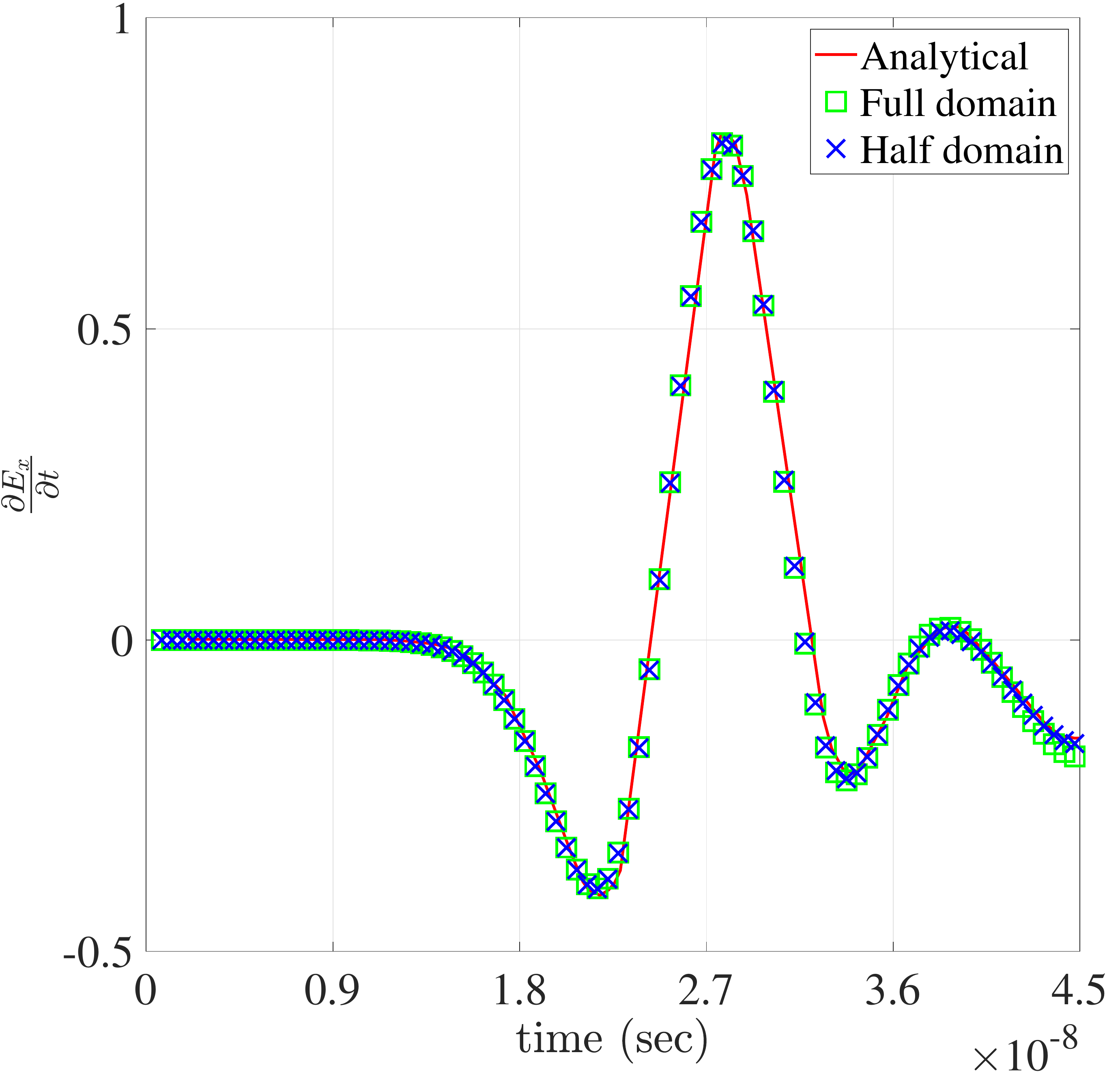}
	\caption{Time variation of the $\parder{E_x}{t}$ at $(x,y,z)=(0.33,-1.03,-0.45)$ for scattering from a conducting sphere.}
\label{scatt_cond_sphere}
\end{figure}

\begin{table}[pos=h!]
	\caption{Mesh and equation details for scattering from a conducting sphere.} \label{table_equations_cond_sphere}
	\begin{center}
		\begin{tabular}{p{0.3\linewidth}p{0.3\linewidth}p{0.2\linewidth}}\toprule
			\multicolumn{1}{l}{Mesh size}  & \multicolumn{1}{l}{half domain}& \multicolumn{1}{l}{full domain~\cite{Nandy2018}}  \\ \midrule
			\multicolumn{1}{l}{$(r\times\theta\times\phi)$} &     $8\times8\times5$ +    &     $8\times8\times10$ \\
			\multicolumn{1}{l}{$(r\times\theta\times t)$} &$8\times16\times1$&      \\ \midrule
			\multicolumn{1}{l}{Number of elements}& 448 & 640 \\ \midrule
			Number of equations &11169    &14946  \\ \bottomrule
		\end{tabular}
	\end{center}
\end{table}
To know the time domain performance of the proposed method on curved objects, we consider the present example where we simulate the transient scattering electric fields from a conducting sphere. An incident wave as stated \text{in}~\cite{Nandy2018} given as 
\begin{equation}
\bE_{\text{inc}} = 2\{t-t_0-\hat{\bk}\cdot(\br-\br_0)/c\}\exp\left[-\frac{\{t-t_0-\hat{\bk}\cdot(\br-\br_0)/c\}^2}{\tau^2}\right]\hat{\bE},
\label{Einc}
\end{equation}
impinges on a conducting sphere. Here, $t_0$ is the temporal reference point, $\br_0$ is the spatial reference point, and $\tau$ is the pulse shape. Following pulse parameters such as $\hat{\bk}=\hat{\bz}$, $\hat{\bE}=\hat{\bx}$, $t_0 = 25.99\ \text{ns}$, $\br_0 = -1.2\hat{\bz}\ \text{m}$ and $\tau = 5.25\ \text{ns}$ are used for simulation. Computational full domain which is truncated at a radius $R_\infty=3.8$ m, is meshed with 640 B27 and W18 elements which results in 14,946 equations in~\cite{Nandy2018}. In the current symmetric boundary condition implementation, the half domain with cylindrical patch is meshed with 448 B27 and W18 elements resulting in 11,169 equations (Table~\ref{table_equations_cond_sphere}). Boundary conditions such as $\bE\cdot\bn=\bzero$ is implemented on the surface of the thin patch as described in section~\ref{method2}, and $\bE\bcross\bn=\bzero$ is applied on the conducting surface. We have taken a time step of $5\times10^{-10} \text{sec}$ in our simulation. From Fig.~\ref{scatt_cond_sphere}, it is evident that $\parder{E_x}{t}$ values at $(0.33,-1.03,-0.45)$ in both full domain and half domain with symmetry, are perfectly matching with the standard benchmark results in~\cite{Jiao2002a}. Around 25$\%$ less unknowns have to be solved using symmetric boundary condition, which clearly represent the computational efficacy of the proposed method. 

\subsection{Electromagnetic scattering from a dielectric sphere} \label{scat_diele_sphere}
\begin{figure}[pos=h!]
 \centering
	\subfloat[$\bE.\bt$ at $(x,y,z)=(-0.04,-0.07,-0.72)$ with $\bt = 0.96\hat{\bx}+0.26\hat{\by}+0.13\hat{\bz}$]{\includegraphics[trim={0cm 0cm 0cm 0cm},width=0.45\columnwidth]{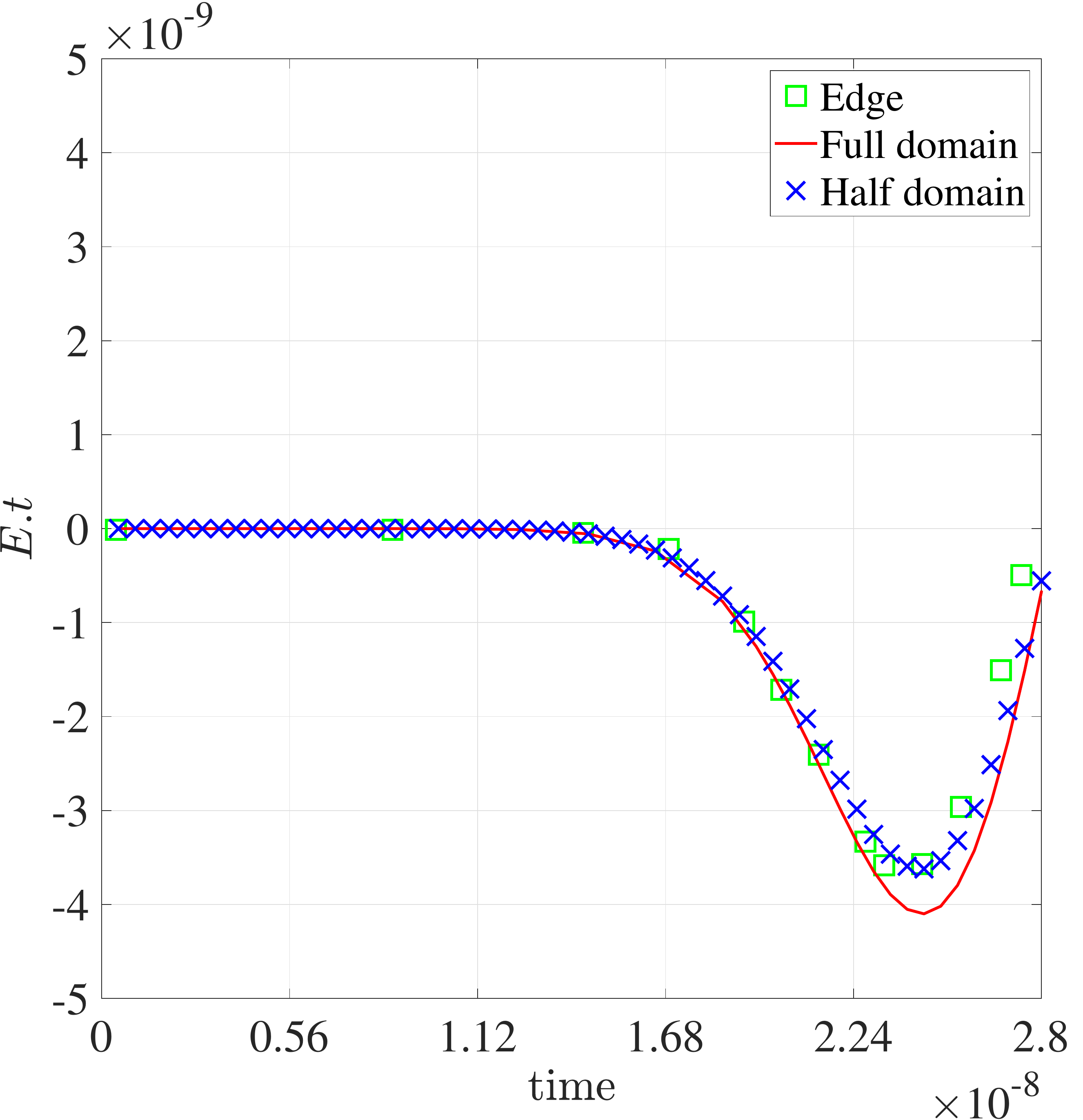}} \hspace{0.2cm} 
	\subfloat[$\bE.\bt$ at $(x,y,z)=(0.05,0.05,-0.96)$ with $\bt = -0.89\hat{\bx}-0.08\hat{\by}+0.44\hat{\bz}$]{\includegraphics[trim={0cm 0cm 0cm 0cm},width=0.45\columnwidth]{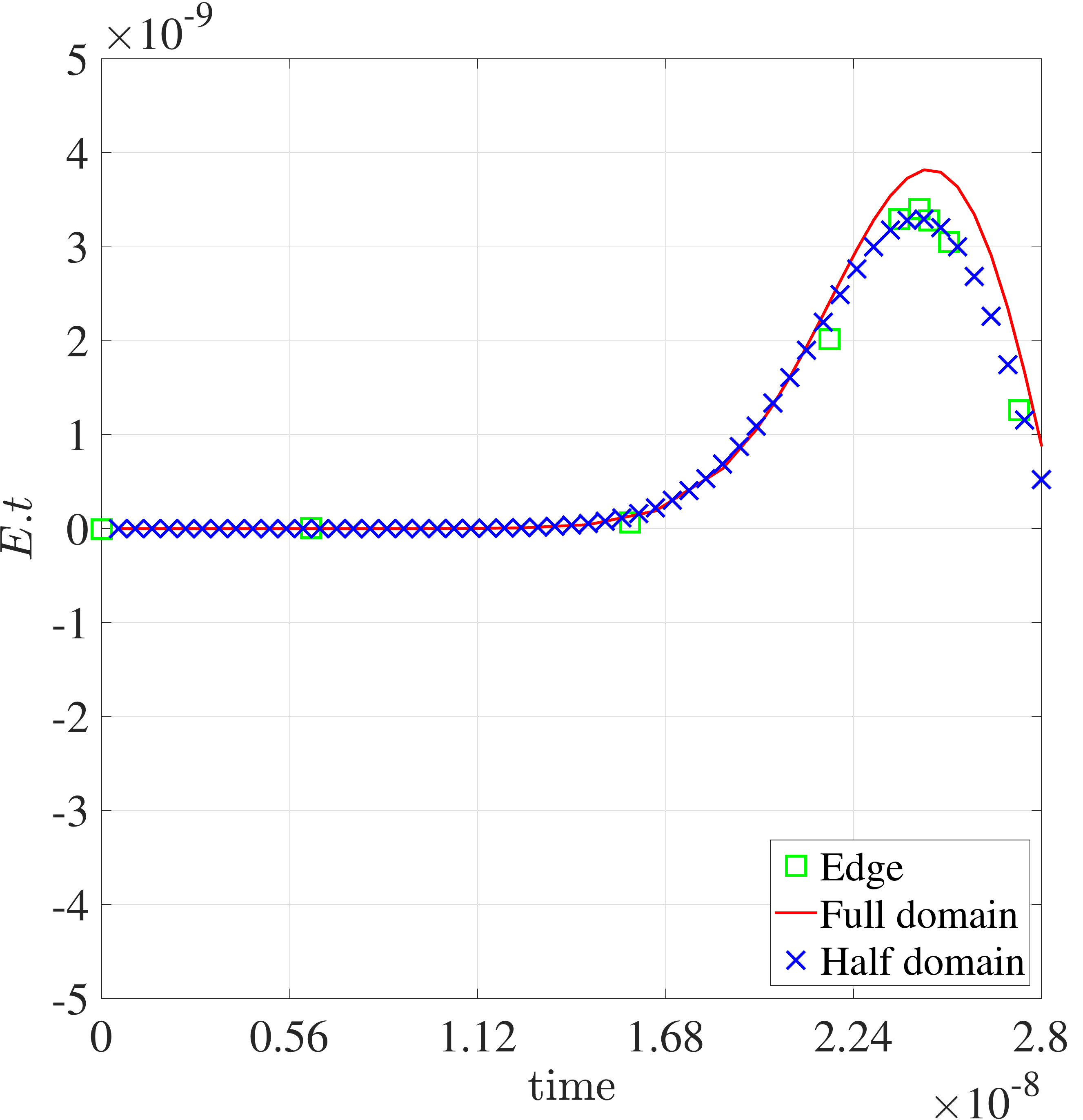}} 
	\caption{Transient variation of the $\bE.\bt$ for scattering from a dielectric sphere.}
\label{dielectric_sphere}
\end{figure}

\begin{table}[pos=h!]
	\caption{Mesh and equation details for scattering from a dielectric sphere.} \label{table_equations_dielectric_sphere}
	\begin{center}
		\begin{tabular}{p{0.3\linewidth}p{0.3\linewidth}p{0.2\linewidth}}\toprule
			\multicolumn{1}{l}{}  & \multicolumn{1}{l}{half domain}& \multicolumn{1}{l}{full domain}  \\ \midrule
			\multicolumn{1}{l}{Mesh size} &     $1536$ (hemispherical domain) +    &     $3702$ \\
			\multicolumn{1}{l}{} &$512$ (patch)&      \\ \midrule
			\multicolumn{1}{l}{Number of elements}& 2048 & 3702 \\ \midrule
			Number of equations &49076    & 70995  \\ \bottomrule
		\end{tabular}
	\end{center}
\end{table}
This example is chosen to test the effectiveness of the proposed method in predicting scattering electric fields from a dielectric body in time domain. On a dielectric sphere of radius $0.5$ m, having $\mu_r=1$ and $\epsilon_r=6.0$, Neumann pulse wave as given by
\begin{equation}
\bE_{\text{inc}} = 2\{t-t_0-\hat{\bk}\cdot(\br-\br_0)/c\}\exp\left[-\frac{\{t-t_0-\hat{\bk}\cdot(\br-\br_0)/c\}^2}{\tau^2}\right]\hat{\bE},
\label{Einc}
\end{equation}
 \text{in}~\cite{Nandy2018} is impinged. The computational domain which is truncated at radius $R_\infty=2 \text{ m}$, is modeled with B27 and W18 elements. Table~\ref{table_equations_dielectric_sphere} gives the numerical analysis details for both full and half domains. Proper symmetric boundary condition is applied with thin cylindrical patch as described in section~\ref{method2}. With a time step of $0.5$ ns, simulations are performed. We have plotted $\bE\cdot\bt$ along time in Fig.~\ref{dielectric_sphere} for full domain, half domain with symmetry and compared it with the edge element results in~\cite{jiao2003time} at two locations: (a) at $(-0.04, -0.07,-0.72)$ with $\bt=0.96\hat{\bx}+0.26\hat{\by}+0.13\hat{\bz}$ and (b) at $(0.05,0.05,-0.96)$ with $\bt=-0.89\hat{\bx}i-0.08\hat{\by}+0.44\hat{\bz}$. Fig.~\ref{dielectric_sphere} shows the efficacy of our proposed method for transient scattering problem with dielectric body. Our proposed method predicts the benchmark electric fields with only 49076 equations which is 30$\%$ less as compared to the number of required equations in full domain analysis.  

\section{Conclusions}
  In the present work, one novel method is proposed to implement symmetric boundary conditions to electromagnetic problems for both harmonic and transient analysis. In nodal framework, potential formulations are followed and it is not very straight forward to implement the symmetric boundary condition in that formulation. In the proposed method, a very thin patch is attached to the symmetry faces of the domain. In that thin patch, two different conditions are required for scalar and vector potentials for proper implementation of the symmetric boundary condition. We have validated the proposed method solving a wide range of problems from radiation to scattering, from spherical to non-spherical domains, from harmonic to transient analysis. For all the problems, we have validated our results against either analytical benchmark results or other benchmark from existing literatures. For harmonic analysis, we have implemented the proposed novel method in amplitude formulation~\cite{harmonic} also. Furthermore, we have presented the computational efficacy of our proposed method as compared to the existing full domain analysis for all the examples.
\section*{Acknowledgments}
The authors gratefully acknowledge the support from SERB, DST under the project IMP/2019/000276, and VSSC, ISRO through MoU No.: ISRO:2020:MOU:NO: 480.

\bibliographystyle{cas-model2-names}
\bibliography{references}				

\end{document}